\documentclass[twocolumn,superscriptaddress,showpacs,showkeys,preprintnumbers,amsmath,amssymb,floatfix,aps,prc,10pt]{revtex4-1}
\usepackage{graphicx}
\usepackage{dcolumn}
\newcolumntype{d}[1]{D{.}{.}{-1}} 
\usepackage{latexsym}
\usepackage{amsmath} 
\usepackage{url}
\usepackage{natbib}
\usepackage{color}
\usepackage{comment}
\usepackage{mathtools}
\usepackage{framed}
\usepackage{diagbox}
\usepackage{multirow}
\usepackage{hyperref}
\usepackage[margin=20mm]{geometry}
\usepackage{siunitx}
\usepackage{subfigure}
\usepackage{makecell}
\newcolumntype{C}[1]{>{\centering\arraybackslash}m{#1}}

\definecolor{forest}{RGB}{0, 153, 51}

\usepackage[normalem]{ulem}
 
\begin{document}

\preprint{Phys. Rev. C}

\title{Measurements of $\gamma_v p \to \pi^+ \pi^-p'$ Cross Sections with the CLAS12 Detector for $Q^2$ from 2.4--8.0~GeV$^2$ and $W$ from 1.4--2.1~GeV}

\newcommand*{\ANL}{Argonne National Laboratory, Argonne, Illinois 60439}
\newcommand*{\ANLindex}{1}
\affiliation{\ANL}
\newcommand*{\SACLAY}{IRFU, CEA, Universit$\acute{e}$ Paris-Saclay, F-91191 Gif-sur-Yvette, France}
\newcommand*{\SACLAYindex}{2}
\affiliation{\SACLAY}
\newcommand*{\CNU}{Christopher Newport University, Newport News, Virginia 23606}
\newcommand*{\CNUindex}{3}
\affiliation{\CNU}
\newcommand*{\UCONN}{University of Connecticut, Storrs, Connecticut 06269}
\newcommand*{\UCONNindex}{4}
\affiliation{\UCONN}
\newcommand*{\DUKE}{Duke University, Durham, North Carolina 27708-0305}
\newcommand*{\DUKEindex}{5}
\affiliation{\DUKE}
\newcommand*{\DUQUESNE}{Duquesne University, Pittsburgh, Pennsylvania 15282}
\newcommand*{\DUQUESNEindex}{6}
\affiliation{\DUQUESNE}
\newcommand*{\FU}{Fairfield University, Fairfield, Connecticut 06824}
\newcommand*{\FUindex}{7}
\affiliation{\FU}
\newcommand*{\FERRARAU}{Universit$\grave{a}$ di Ferrara, 44121 Ferrara, Italy}
\newcommand*{\FERRARAUindex}{8}
\affiliation{\FERRARAU}
\newcommand*{\FIU}{Florida International University, Miami, Florida 33199}
\newcommand*{\FIUindex}{9}
\affiliation{\FIU}
\newcommand*{\Genova}{Universit$\grave{a}$ di Genova, 16146 Genova, Italy}
\newcommand*{\Genovaindex}{10}
\affiliation{\Genova}
\newcommand*{\GWUI}{The George Washington University, Washington, DC 20052}
\newcommand*{\GWUIindex}{11}
\affiliation{\GWUI}
\newcommand*{\GSIFFN}{GSI Helmholtzzentrum fur Schwerionenforschung GmbH, D-64291 Darmstadt, Germany}
\newcommand*{\GSIFFNindex}{12}
\affiliation{\GSIFFN}
\newcommand*{\ORSAY}{Universit$\acute{e}$ Paris-Saclay, CNRS/IN2P3, IJCLab, 91405 Orsay, France}
\newcommand*{\ORSAYindex}{13}
\affiliation{\ORSAY}
\newcommand*{\INFNCAT}{INFN, Sezione di Catania, 95123 Catania, Italy}
\newcommand*{\INFNCATindex}{14}
\affiliation{\INFNCAT}
\newcommand*{\INFNFE}{INFN, Sezione di Ferrara, 44100 Ferrara, Italy}
\newcommand*{\INFNFEindex}{15}
\affiliation{\INFNFE}
\newcommand*{\INFNFR}{INFN, Laboratori Nazionali di Frascati, 00044 Frascati, Italy}
\newcommand*{\INFNFRindex}{16}
\affiliation{\INFNFR}
\newcommand*{\INFNGE}{INFN, Sezione di Genova, 16146 Genova, Italy}
\newcommand*{\INFNGEindex}{17}
\affiliation{\INFNGE}
\newcommand*{\INFNRO}{INFN, Sezione di Roma Tor Vergata, 00133 Rome, Italy}
\newcommand*{\INFNROindex}{18}
\affiliation{\INFNRO}
\newcommand*{\INFNTUR}{INFN, Sezione di Torino, 10125 Torino, Italy}
\newcommand*{\INFNTURindex}{19}
\affiliation{\INFNTUR}
\newcommand*{\INFNPAV}{INFN, Sezione di Pavia, 27100 Pavia, Italy}
\newcommand*{\INFNPAVindex}{20}
\affiliation{\INFNPAV}
\newcommand*{\JMU}{James Madison University, Harrisonburg, Virginia 22807}
\newcommand*{\JMUindex}{21}
\affiliation{\JMU}
\newcommand*{\KNU}{Kyungpook National University, Daegu 41566, Republic of Korea}
\newcommand*{\KNUindex}{22}
\affiliation{\KNU}
\newcommand*{\LAMAR}{Lamar University, Beaumont, Texas 77710}
\newcommand*{\LAMARindex}{23}
\affiliation{\LAMAR}
\newcommand*{\MIT}{Massachusetts Institute of Technology, Cambridge, Massachusetts 02139-4307}
\newcommand*{\MITindex}{24}
\affiliation{\MIT}
\newcommand*{\MISS}{Mississippi State University, Mississippi State, Mississippi 39762-5167}
\newcommand*{\MISSindex}{25}
\affiliation{\MISS}
\newcommand*{\MSU}{Skobeltsyn Institute of Nuclear Physics and Physics Department, Lomonosov Moscow State University, 119234 Moscow, Russia}
\newcommand*{\MSUindex}{26}
\affiliation{\MSU}
\newcommand*{\UNH}{University of New Hampshire, Durham, New Hampshire 03824-3568}
\newcommand*{\UNHindex}{27}
\affiliation{\UNH}
\newcommand*{\NMSU}{New Mexico State University, Las Cruces, New Mexico 88003}
\newcommand*{\NMSUindex}{28}
\affiliation{\NMSU}
\newcommand*{\OHIOU}{Ohio University, Athens, Ohio 45701}
\newcommand*{\OHIOUindex}{29}
\affiliation{\OHIOU}
\newcommand*{\ODU}{Old Dominion University, Norfolk, Virginia 23529}
\newcommand*{\ODUindex}{30}
\affiliation{\ODU}
\newcommand*{\JLUGiessen}{II Physikalisches Institut der Universitaet Giessen, 35392 Giessen, Germany}
\newcommand*{\JLUGiessenindex}{31}
\affiliation{\JLUGiessen}
\newcommand*{\URICH}{University of Richmond, Richmond, Virginia 23173}
\newcommand*{\URICHindex}{32}
\affiliation{\URICH}
\newcommand*{\ROMAII}{Universi$\grave{a}$ di Roma Tor Vergata, 00133 Rome Italy}
\newcommand*{\ROMAIIindex}{33}
\affiliation{\ROMAII}
\newcommand*{\SDU}{Shandong University, Qingdao, Shandong 266237, China}
\newcommand*{\SDUindex}{34}
\affiliation{\SDU}
\newcommand*{\SCAROLINA}{University of South Carolina, Columbia, South Carolina 29208}
\newcommand*{\SCAROLINAindex}{35}
\affiliation{\SCAROLINA}
\newcommand*{\TEMPLE}{Temple University, Philadelphia, Pennsylvania 19122}
\newcommand*{\TEMPLEindex}{36}
\affiliation{\TEMPLE}
\newcommand*{\JLAB}{Thomas Jefferson National Accelerator Facility, Newport News, Virginia 23606}
\newcommand*{\JLABindex}{37}
\affiliation{\JLAB}
\newcommand*{\ULS}{Universidad de La Serena, Avda. Juan Cisternas 1200, La Serena, Chile}
\newcommand*{\ULSindex}{38}
\affiliation{\ULS}
\newcommand*{\UTFSM}{Universidad T\'{e}cnica Federico Santa Mar\'{i}a, Casilla 110-V Valpara\'{i}so, Chile}
\newcommand*{\UTFSMindex}{39}
\affiliation{\UTFSM}
\newcommand*{\BRESCIA}{Universit$\grave{a}$ degli Studi di Brescia, 25123 Brescia, Italy}
\newcommand*{\BRESCIAindex}{40}
\affiliation{\BRESCIA}
\newcommand*{\UCR}{University of California Riverside, Riverside, California 92521}
\newcommand*{\UCRindex}{41}
\affiliation{\UCR}
\newcommand*{\GLASGOW}{University of Glasgow, Glasgow G12 8QQ, United Kingdom}
\newcommand*{\GLASGOWindex}{42}
\affiliation{\GLASGOW}
\newcommand*{\YORK}{University of York, York YO10 5DD, United Kingdom}
\newcommand*{\YORKindex}{43}
\affiliation{\YORK}
\newcommand*{\VIRGINIA}{University of Virginia, Charlottesville, Virginia 22901}
\newcommand*{\VIRGINIAindex}{44}
\affiliation{\VIRGINIA}
\newcommand*{\WM}{College of William and Mary, Williamsburg, Virginia 23187-8795}
\newcommand*{\WMindex}{45}
\affiliation{\WM}
\newcommand*{\YEREVAN}{Yerevan Physics Institute, 375036 Yerevan, Armenia}
\newcommand*{\YEREVANindex}{46}
\affiliation{\YEREVAN}

\newcommand*{\NOWODU}{Old Dominion University, Norfolk, Virginia 23529}
\newcommand*{\NOWUCONN}{University of Connecticut, Storrs, Connecticut 06269}

\newcommand*{\NOWPARISSACLAY}{DOTA, ONERA, Université Paris-Saclay, 91120 Palaiseau, France}

\newcommand*{\NOWJNRDRUSSIA}{Joint Institute for Nuclear Research, 141980 Dubna, Russia}
\newcommand*{\NOWLBNL}{Lawrence Berkeley National Laboratory, California, 94720}

\author {K.~Neupane} 
\affiliation{\JLAB}
\affiliation{\SCAROLINA}
\author {R.W.~Gothe} 
\affiliation{\SCAROLINA}
\author {D.S.~Carman}
\affiliation{\JLAB}
\author {V.I.~Mokeev} 
\affiliation{\SCAROLINA}
\affiliation{\JLAB}
\author {A.G.~Acar} 
\affiliation{\YORK}
\author {P.~Achenbach} 
\affiliation{\CNU}
\author {J.S.~Alvarado} 
\affiliation{\ORSAY}
\author {W.R.~Armstrong} 
\affiliation{\ANL}
\author {H.~Avakian} 
\affiliation{\JLAB}
\author {N.A.~Baltzell} 
\affiliation{\JLAB}
% ...

\collaboration{The CLAS Collaboration}
\noaffiliation

\date{\today}

\begin{abstract}
This paper reports exclusive cross sections for the $ep \to e'\pi^+\pi^-p'$ reaction using the CLAS12 detector at Jefferson Laboratory. The extractions of fully integrated and nine single-differential cross sections are presented for the first time for photon virtualities $Q^2$ from 2.4 to 8.0~GeV$^2$ and center-of-mass energies $W$ from 1.4 to 2.1~GeV, which covers a large part of the nucleon resonance region. These data considerably extend the kinematic reach of previous measurements from CLAS that covered $Q^2$ up to 5.0~GeV$^2$. Exclusive $\gamma_v p \to \pi^+ \pi^-p'$ cross section measurements are of particular importance for the extraction of $\gamma_vpN^*$ resonance electrocouplings across the $N^*$ spectrum, especially in the mass range above 1.6~GeV where several resonances decay preferentially to the $\pi\pi N$ final states. The electrocouplings with extended $Q^2$ coverage expected from these new data will enable an improved understanding of the emergence of $N^*$ mass and structure in the transition from the strongly coupled toward the perturbative QCD regime.
\end{abstract}

\maketitle


\section{Introduction}
\label{sec:intro}

Studies of the structure of excited nucleon states ($N^*$s) for prominent resonances across the full $N^*$ spectrum through the evolution of their $\gamma_vpN^*$ electroexcitation amplitudes (commonly referred to as $\gamma_vpN^*$ electrocouplings) as a function of the squared four‑momentum of the virtual photon, $Q^2$, provide a powerful means to explore how quantum chromodynamics (QCD) generates these states and their internal structure as bound systems of quarks and gluons \cite{Burkert:2025coj,Mokeev:2022xfo,Aznauryan:2011qj,Burkert:2019bhp}. These studies offer unique information on many facets of strong‑interaction dynamics in the regime where the QCD running coupling is large and close to unity, the so‑called strong‑QCD (sQCD) regime, as manifested in the generation of $N^*$ states with different masses, spin‑parity quantum numbers, and structural features~\cite{Cheng:2025sdp,Burkert:2025coj,Mokeev:2022xfo,Skorodumina:2015ccu}. 

Experiments performed during the 6-GeV era with the CLAS detector~\cite{CLAS:2003umf} at Jefferson Laboratory (JLab) have provided the dominant share of the world’s data on differential and integrated cross sections, as well as various polarization asymmetries, for most exclusive meson electroproduction channels in the resonance region of hadronic invariant masses $W < 2.0$~GeV and $Q^2 < 5.0$~GeV$^2$ \cite{Burkert:2025coj,Aznauryan:2011qj,Mokeev:2022xfo}. Independent analyses of the data from these different channels \cite{Aznauryan:2009mx,Park:2014yea,Denizli:2007tq,Mokeev:2008iw,Mokeev:2012vsa,Mokeev:2023zhq} have yielded information on the $\gamma_vpN^*$ electrocouplings of most $N^*$ states  across the mass range up to 1.8~GeV for $Q^2 < 5.0$~GeV$^2$~\cite{HillerBlin:2019jgp}. Recently, the first results on the $\gamma_vpN^*$ electrocouplings from a global multichannel analysis of pseudoscalar meson-baryon photo‑, electro‑, and hadroproduction data have become available, obtained within an advanced coupled‑channel approach developed by the J{\"u}lich-Bonn-Washington (JBW) Collaboration~\cite{Wang:2024byt}.

The successful description of the pion and nucleon elastic electromagnetic form factors, together with the $\gamma_vpN^*$ electrocouplings of the $\Delta(1232)3/2^+$, $N(1440)1/2^+$, and $\Delta(1600)3/2^+$ resonances, each characterized by a markedly different internal structure, has been achieved within the continuum Schwinger method (CSM)~\cite{Achenbach:2025kfx, Carman:2023zke, Ding:2022ows} using the same momentum dependence of the dressed‑quark mass derived from the QCD Lagrangian. This success clearly demonstrates the capability of the CSM approach to provide deep insights into the emergence of more than 98\% of hadron mass by analyzing the experimental results on the $Q^2$-evolution of $\gamma_vpN^*$ electrocouplings. These studies directly address one of the most challenging open problems of the Standard Model: understanding the nature of the dominant component of the visible mass in the Universe.

Experiments of the 6-GeV era on meson electroproduction with the CLAS detector were limited to $Q^2 < 5.0$~GeV$^2$, allowing for the exploration of only a limited range of distances in which $\approx 30\%$ of hadron mass is generated. In this domain, $N^*$ structure is determined by a complex interplay between an inner core of three dressed quarks and an external meson-baryon cloud~\cite{Burkert:2019bhp,Burkert:2025coj,Aznauryan:2012ba,Suzuki:2010yn}, with a gradual transition to quark-core dominance across the range $2.0 < Q^2 < 5.0$~GeV$^2$. 

Studies of $\pi^+\pi^-p$ electroproduction from CLAS data have also advanced knowledge on the $N^*$ spectrum. The combined analysis of $\pi^+\pi^-p$ photo‑ and electroproduction data led to the discovery of a new $N'(1720)3/2^+$ baryon state~\cite{Mokeev:2020hhu}. Currently, it is the only known new $N^*$ state for which information on the $Q^2$ evolution of the $\gamma_vpN^*$ electrocouplings has become available. This information provides a promising opportunity to shed light on the peculiar structural features of new $N^*$ states that have made their observation so elusive for decades.  

The CLAS detector of the JLab 6-GeV era was decommissioned in 2012. During the period from 2012 to 2017, the JLab CEBAF electron accelerator was upgraded to support 12-GeV operations, and CLAS was replaced by the CLAS12 detector~\cite{Burkert:2020akg}. Currently, CLAS12 is the only facility capable of providing information to map out the evolution of the $\gamma_v p N^*$ electrocouplings across the largely unexplored range of $Q^2$ up to 10~GeV$^2$, representing the highest $Q^2$ values ever achieved for meson electroproduction over the $N^*$ resonance region. Measurements with the CLAS12 detector are focused on extending the experimental results from independent and combined analyses of $\pi^0 p$, $\pi^+ n$, $K^+ \Lambda$, $K^+ \Sigma^0$, and $\pi^+ \pi^- p$ electroproduction off protons \cite{Proceedings:2020fyd}. 

The first experimental results from CLAS12 on inclusive $(e,e'X)$ cross sections in the region $W < 2.5$~GeV for $Q^2 = 2.4-10.0$~GeV$^2$ revealed the presence of resonance-like structures in the first, second, and third resonance regions across the entire $Q^2$ range covered by the measurements \cite{CLAS:2025zup}. These findings offer a promising prospect for extracting the $\gamma_vpN^*$ electrocouplings over this kinematic range from the observables of the exclusive meson electroproduction channels.

The anticipated results on the evolution of the $\gamma_v p N^*$ electrocouplings across the $Q^2$ range up to 10.0~GeV$^2$ will allow exploration of the range of distances where up to 50\% of hadron mass is generated. In this regime, the core of three dressed quarks is the major contributor to the structure of the $N^*$ states. Studies of the $Q^2$ evolution of the $\gamma_v p N^*$ electrocouplings from the CLAS12 experimental data will shed light on the evolution of the quark-gluon component of $N^*$ structure over the range of the distances where the transition from the strongly coupled (sQCD) toward the perturbative QCD (pQCD) regime takes place~\cite{Achenbach:2025kfx,Ding:2022ows,Cheng:2025sdp}.

The $\pi^0 p$, $\pi^+ n$, and $\pi^+ \pi^- p$ electroproduction channels are the major contributors in the nucleon resonance region. The combination of results from studies of these channels allows for the exploration of the $\gamma_v p N^*$ electrocouplings for almost all prominent excited states of the nucleon. The low-lying $N^*$ states across the mass range below 1.6~GeV decay preferentially to $\pi N$ final states but still have substantial branching fractions, around 40\%, for decays to $\pi\pi N$. This allows for the extraction of the electrocouplings of low-lying $N^*$ states from studies of both $\pi N$ and $\pi^+ \pi^- p$ electroproduction. In the mass range above 1.6~GeV, several $N^*$ states decay predominantly into $\pi\pi N$, making the $ep \rightarrow e'\pi^+\pi^-p'$ channel the primary source of information on their structure.

This paper reports the first results from CLAS12 on the $\pi^+ \pi^- p$ electroproduction cross sections in the region of $W < 2.1$~GeV for $Q^2 = 2.4-8.0$~GeV$^2$. These new data extend the existing $\pi^+\pi^-p$ electroproduction data available from CLAS over the range of $W < 2.0$~GeV and $Q^2 = 0.4-5.0$~GeV$^2$ \cite{CLAS:2002xbv,CLAS:2017fja,CLAS:2018fon,Trivedi:2018rgo}. In each $(W,Q^2)$ bin, nine independent one-fold differential cross sections have been extracted. They will serve as the first available experimental input to update the JLab-Moscow State University (JM) reaction model, which has been successfully employed for the extraction of the $\gamma_v p N^*$ electrocouplings from measurements with CLAS at $Q^2 < 5.0$~GeV$^2$, to establish the reaction mechanisms for $\pi^+ \pi^- p$ electroproduction over the expanded $Q^2$ range of the CLAS12 data.

The organization for the remainder of this paper is as follows. In Section~\ref{sec:formalism} the cross sections extracted from this analysis of $\pi^+\pi^-p$ electroproduction are defined along with the relevant kinematic variables. Section~\ref{sec:analysis} describes the experiment, data binning, event selection, and analysis procedures, including a newly developed and essential reconstruction efficiency correction method. Section~\ref{sec:mc} details the detector Monte Carlo (MC) simulations and the event generator, as well as the procedure for filling the holes from the zero-acceptance regions of the detector. The statistical and systematic uncertainties associated with the cross section measurement are discussed in Section~\ref{sec:uncertainties}. Section~\ref{sec:results} presents the cross sections and discusses the prospects for extension of the information on the $\gamma_vpN^*$ electrocouplings from this data across the still almost unexplored range of $Q^2$=5.0--8.0 GeV$^2$. Section~\ref{sec:conclusions} provides a final summary and conclusions.

\section{Cross Section Formalism}
\label{sec:formalism}

The kinematics of the $ep \to e' \pi^+ \pi^- p'$ reaction is fully determined by the invariant mass $W$, the photon virtuality $Q^2$, and by five additional variables that define the four-momenta of the $\pi^+ \pi^- p$ final-state hadrons~\cite{Fedotov:2008aa,CLAS:2017fja}. The four-momenta of the three hadrons in the final state are described by a total of 12 variables. Energy–momentum conservation imposes four constraints, while the on-shell conditions for the three final state hadrons impose three additional constraints. Hence, the three-body final-state kinematics is fully determined by only five independent hadronic variables. As in inclusive scattering, $W$ and $Q^2$ fully determine the remaining kinematics of the initial target proton and virtual photon (or scattered electron) state. Consequently, the measured $ep \to e' \pi^+ \pi^- p'$ cross sections are seven-fold differential, while the virtual photon $\gamma_v p \to \pi^+ \pi^- p'$ cross sections within any given $(W,Q^2)$ bin are five-fold differential. In our notation we use $d^5\tau$ to denote the differential in these five independent kinematic variables.

These cross sections can be obtained in three sets of hadronic variables depending on various choices for the first ($h_1$), second ($h_2$), and third ($h_3$) final state hadrons. In general, the invariant masses of the first pair of hadrons, $M_{h_1h_2}$, the second pair of hadrons, $M_{h_2h_3}$, and the angles of the first hadron are included in each set of hadronic variables, as described below. All frame-dependent variables should be in the center-of-mass (CM) frame for the final state hadrons. See Ref.~\cite{Fedotov:2008aa,CLAS:2017fja,CLAS:2018fon,Trivedi:2018rgo,CLAS:2023mfc,iuliia_th} for more details.

\begin{figure}[htbp]
\begin{center}
\includegraphics[width=1.0\columnwidth]{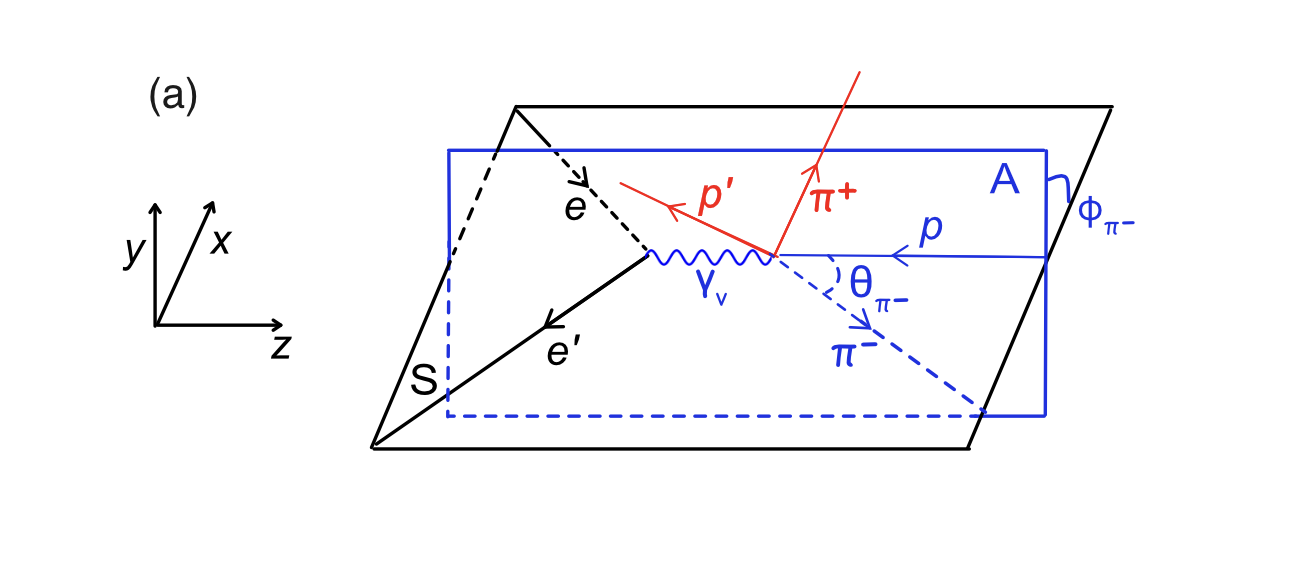}
\includegraphics[width=1.0\columnwidth]{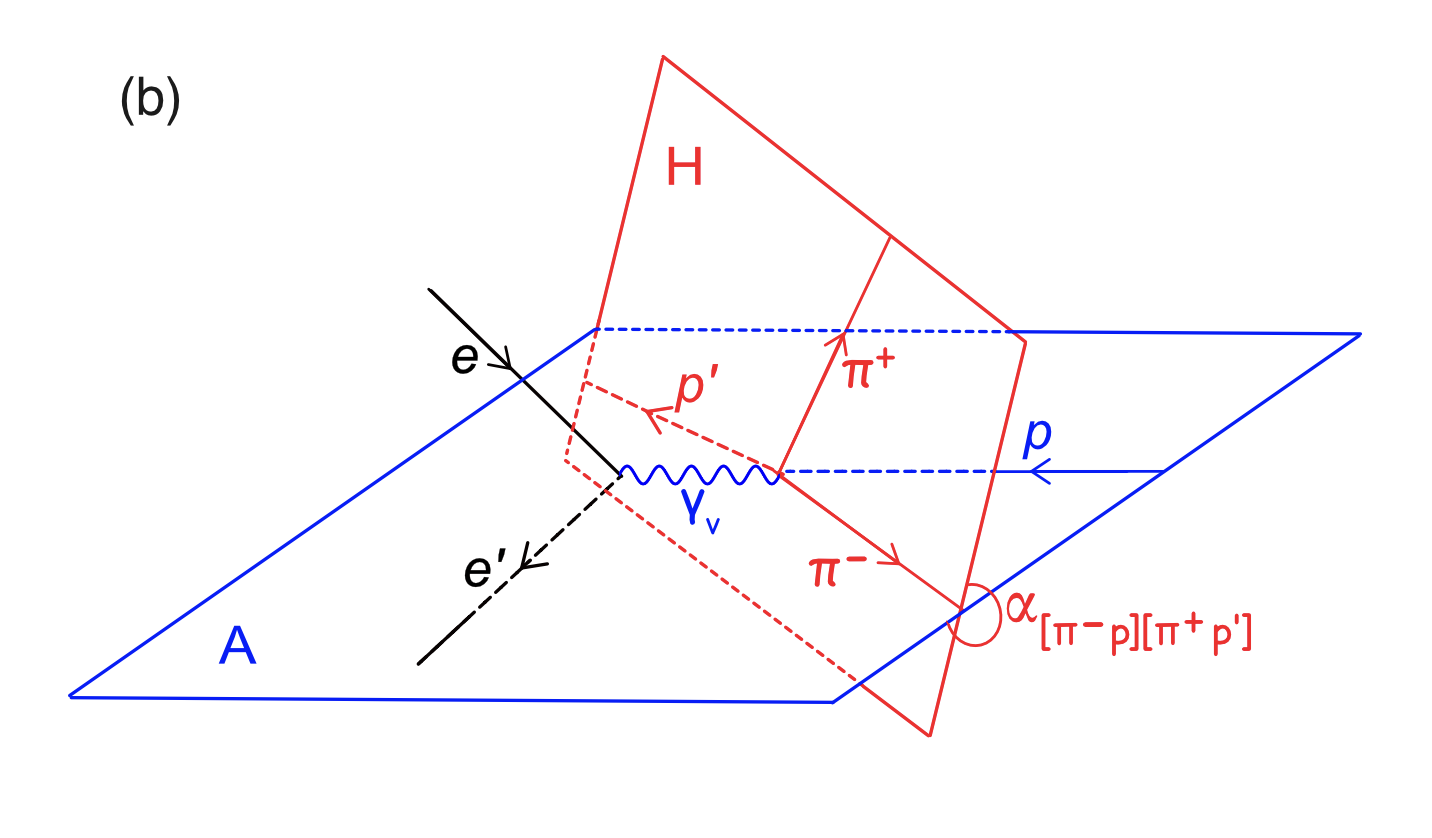}
\caption{Kinematic variables for the description of the reaction $\gamma_v p \to  \pi^+ \pi^- p'$ in the CM frame of the final state hadrons for Set 1. Panel (a) shows the $\pi^-$ polar and azimuthal angles $\theta_{\pi^-}$ and $\phi_{\pi^-}$. Plane S represents the electron scattering plane. The $z$-axis is directed along the virtual photon $\gamma_v$, while the $x$-axis, orthogonal to the photon three-momentum, is located in the electron scattering plane S, and the $y$-axis forms a right-handed coordinate system. Plane A is defined by the three-momenta of the initial state proton and the final state $\pi^-$. Panel (b) shows the angle $\alpha_{[\pi^-p][\pi^+p']}$ between the two hadronic planes A and H or the plane H rotation angle around the axis aligned along the three-momentum of the final state $\pi^-$. Plane H is defined by the three-momenta of the final state $\pi^+$ and $p'$.}
\label{fig:kinematics}
\end{center}
\end{figure}

\begin{itemize}

\item[]{Set 1:} Invariant mass of the $\pi^+\pi^-$ pair $M_{\pi^+\pi^-}$, invariant mass of the $\pi^+p'$ pair $M_{\pi^+p'}$, $\pi^-$ spherical angles $\theta_{\pi^-}$ and $\phi_{\pi^-}$, and the angle between the plane defined by the three-momenta of the final state $\pi^+$ and $p'$ and the plane defined by the three-momenta of the final state $\pi^-$ and the initial state proton $\alpha_{[\pi^-p][\pi^+p']}$, shown in Fig.~\ref{fig:kinematics}.

\item[]{Set 2:} Invariant mass of the $p'\pi^+$ pair $M_{p'\pi^+}$, invariant mass of the $\pi^+\pi^-$ pair $M_{\pi^+\pi^-}$, spherical angles of the proton $\theta_{p'}$ and $\phi_{p'}$, and the angle between the plane defined by the three-momenta of the final state $\pi^+$ and $\pi^-$ and the plane defined by the three-momenta of the initial and final state protons $\alpha_{[\pi^+\pi^-][pp']}$.

\item[]{Set 3:} Invariant mass of the $\pi^+\pi^-$ pair $M_{\pi^+\pi^-}$, invariant mass of the $\pi^-p'$ pair $M_{\pi^-p'}$, $\pi^+$ spherical angles $\theta_{\pi^+}$ and $\phi_{\pi^+}$, and the angle between the plane defined by the momenta of the final state $\pi^-$ and proton and the plane defined by the three-momenta of the final state $\pi^+$ and  the initial state proton $\alpha_{[\pi^+p][\pi^-p']}$.
\end{itemize}

In the single-photon exchange approximation, the five-dimensional (5D) hadronic or virtual photon cross section in a given bin of $Q^2$ and $W$ is related to the seven-dimensional (7D) electron scattering cross section by

\begin{multline}
    \frac{d^5\sigma}{dM_{h_1 h_2}dM_{h_2 h_3}d\Omega_{h_1} d\alpha_{[h_1 p][h_2 h_3]}} =\\
    \frac{1}{\Gamma_\upsilon}\frac{d^7\sigma}{dW dQ^2 dM_{h_1h_2}dM_{h_2h_3}d\Omega_{h_1} d\alpha_{[h_1p][h_2 h_3]}},
\end{multline}
\noindent
where the virtual photon flux~\cite{Akerlof:1967zza} is
\begin{equation}
    \Gamma_v = \frac{\alpha}{4\pi} \frac{1}{E_b^2 M_p^2} \frac{W(W^2 - M_p^2)}  {(1 - \epsilon) Q^2}.
\end{equation}
\noindent
Here $\alpha$ is the fine structure constant, $E_b$ is the electron beam energy, $M_p$ is the proton mass, and $\epsilon$ is the transverse polarization of the virtual photon \cite{Akerlof:1967zza}, given by
\begin{equation}
    \epsilon = \left(1+2\left(1+\frac{\nu^2}{Q^2}\right) \tan^2\!\left(\frac{\theta_{e'}}{2}\!\right)\!\right)^{-1},
\end{equation}
\noindent
where $\theta_{e'}$ is the scattered electron polar angle and $\nu$ is the difference between the energy of the beam and the scattered electron, both in the lab frame.

In the context of two-pion electroproduction, it becomes especially crucial to have enough measured and simulated data. Even for $(W,Q^2)$ bins with exceptionally high statistics, achieving a 5D virtual photon cross section with reasonable accuracy is virtually impossible owing to the enormous number of 7D-cells on the order of $10^6$ to $10^7$. As a result, nine independent single-differential virtual photon cross sections are extracted within each $(W,Q^2)$ bin. This is accomplished by integrating the five-fold differential cross sections over four hadronic variables. An example of a single-differential cross section for the invariant mass of the first pair of hadrons is given by

\begin{equation}
    \frac{d\sigma}{dM_{h_1h_2}}=\int\frac{d^5\sigma}{d^5\tau} dM_{h_2h_3}d\Omega_{h_1}d\alpha_{[h_1 p][h_2 h_3]}.
    \label{equ:1-d_cross_section}
\end{equation}
Similarly, all other single-differential cross sections can be extracted by integrating over the remaining four hadronic variables. 

From the three sets of hadronic variables, nine independent single-differential cross sections were extracted. These nine single-differential cross sections are used by the JM model~\cite{Mokeev:2015lda} to extract the reaction amplitudes. These cross sections are of particular interest for identifying the meson-baryon channels that contribute to $\pi^+\pi^-p$ electroproduction. For example, the invariant mass $M_{\pi^+p}$ distribution shows a characteristic peak corresponding to the $\Delta^{++}$ resonance that elucidates the contribution of the $\pi^-\Delta^{++}$ channel. The $\theta$ distributions help to separate the contributions of $s$-channel resonances and non-resonant amplitudes. The $\alpha$ angle distributions are sensitive to the density matrix of the $\pi^+ p'$, $\pi^+\pi^-$, and $\pi^-p'$ final states. Owing to parity conservation in the strong and electromagnetic interactions, the $\alpha$ angle distributions have to be left-right symmetric with respect to 180$^\circ$.
\section{Experiment Description and Data Analysis}
\label{sec:analysis}

\subsection{Detector and Dataset Information}

The large acceptance CLAS12 spectrometer~\cite{Burkert:2020akg} was installed in Hall~B as part of the JLab 12-GeV upgrade project with the physics program beginning in 2018. CLAS12 consists of both a Forward Detector (FD) system and a Central Detector (CD) system. 

The FD is built around a superconducting torus magnet that divides the acceptance into six 60$^\circ$-wide azimuthal sectors. The torus produces a field primarily in the azimuthal direction with an $\int \!B d\ell$ field strength varying from 2.78~Tm at 5$^\circ$ to 0.54~Tm at 40$^\circ$. The electric charge of a particle will dictate its curvature as it traverses the toroidal field. The field deflects particles along the polar angle, either making their curvature ``inbending" or ``outbending" based on the field polarity. Reconstruction algorithms take the track curvature into account to assign a charge. The FD consists of multiple sets of drift chambers (DC) \cite{Mestayer:2020saf} for charged particle tracking installed before, within, and after the torus magnetic field. Downstream of the DC, the spectrometer includes a forward time-of-flight system (FTOF) for precise timing measurements of charged particles~\cite{Carman:2020fsv} and a sampling electromagnetic calorimeter (ECAL) for identification of electrons, photons, and neutrons~\cite{Asryan:2020iqj}. The FD also includes a high-threshold Cherenkov detector (HTCC) ~\cite{Sharabian:2020whm} upstream of the DC that is used as part of the electron trigger to separate electrons from pions. The FD covers polar angles $\theta$ from 5$^\circ$--35$^\circ$. 

\begin{figure}[htbp]
  \centering
  \includegraphics[width=1.0\columnwidth]{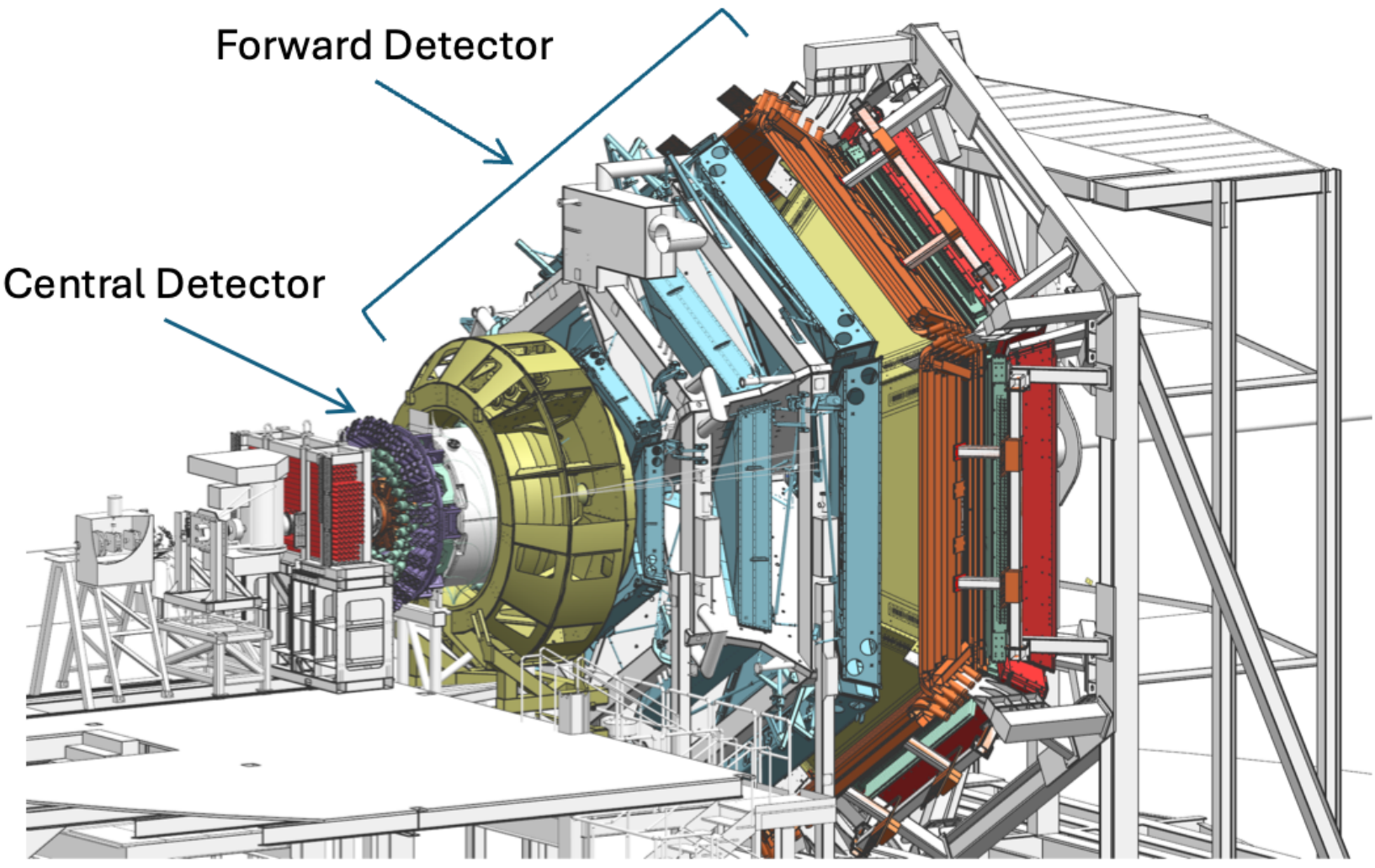}
  \caption{Model of the CLAS12 spectrometer in Hall~B at JLab. The electron beam is incident from the left. The detector system is $\approx 20$~m long along the beam $z$ axis. See Ref.~\cite{Burkert:2020akg} for details.}
  \label{clas12-model}
\end{figure}

The CD system is contained with a 5-T superconducting solenoid magnet and covers polar angles from 35$^\circ$--125$^\circ$. It consists of a multi-layer central vertex tracker (CVT) system divided effectively into three azimuthal sectors surrounding the target and beamline \cite{Antonioli:2020ylv,Acker:2020qkv} and a central time-of-flight system (CTOF) for charged particle identification \cite{Carman:2020yma}. Figure~\ref{clas12-model} shows a model representation of CLAS12. 

The data employed for this work were collected as part of the Run Group~A (RG-A) experiment that took place in the fall of 2018 with production data-taking at beam currents of 45~nA, 50~nA, and 55~nA, amounting to 72\%, 10\%, and 18\% of the collected charge, respectively. Data were acquired at a beam energy of 10.604~GeV using a 5-cm-long liquid-hydrogen target at an average beam-target luminosity of $\sim 5\times10^{34}$~cm$^{-2}$s$^{-1}$. The torus magnet was set to its maximum field strength to optimize the momentum resolution for charged particles and its polarity was set to bend negatively charged particles toward the beamline. The data acquisition system~\cite{Boyarinov:2020yry} recorded data at typical rates of 15~kHz and 500~MB/s with a live-time greater than 90\%.

The readout of CLAS12 was based on a data acquisition trigger defined by a candidate electron track through the drift chambers into the ECAL. The trigger was based on pre-defined DC trajectories that matched to a cluster in the HTCC with a threshold of two photoelectrons, a cluster in the ECAL with a minimum deposited energy of 300~MeV, and a minimum energy deposited in specific ECAL sub-layers (called PCAL, ECin, ECout) in the same CLAS12 sector~\cite{Raydo:2020lxn}. This trigger configuration provided an efficiency exceeding 99\%.

\subsection{Data Binning}
\label{binning}

\begin{figure}[ht]
  \centering
  \includegraphics[width=0.95\columnwidth]{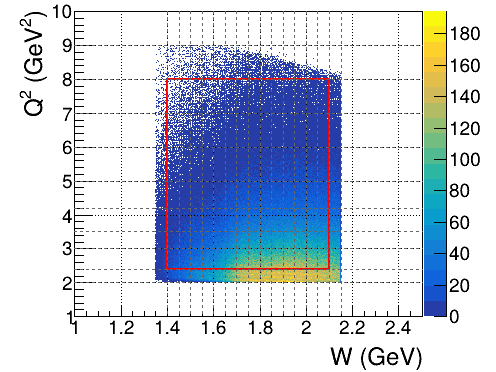}
  \caption{$Q^2$ vs. $W$ distribution for this analysis of the RG-A fall 2018 inbending two-pion dataset. The red box represents the selected region for the analysis and the grid lines represent the bin edges.}
  \label{fig:w_q2_coverage}
\end{figure}

The CLAS12 detector effectively covers nearly the full $4\pi$ kinematic angular range over a wide range of $W$ and $Q^2$. This analysis accesses the full nucleon resonance region and $Q^2$ up to 8~GeV$^2$. Figure~\ref{fig:w_q2_coverage} shows the $Q^2$ and $W$ kinematic region covered by this analysis, along with the binning indicated by the horizontal and vertical grid lines. The bin sizes for $W$ and $Q^2$ reflect a balance between precision, which favors smaller bins, and the available statistics and detector resolution, which favor larger bins. The $W$ bins span the range from 1.4 to 2.1~GeV with a bin width of 50~MeV, consistent with the CLAS12 resolution \cite{CLAS:2025zup}. The width of the $Q^2$-bins range from 0.5~GeV$^2$ to 1.0~GeV$^2$ to compensate for the rapid fall-off of the cross section with increasing $Q^2$. The seven $Q^2$ bins defined for this analysis are: [2.4:3.0], [3.0:3.5], [3.5:4.2], [4.2:5.0], [5.0:6.0], [6.0:7.0], and [7.0:8.0]~GeV$^2$.

For the hadronic variables, this analysis employs 14 bins for the di-hadron invariant mass distributions within their respective ranges, ten bins for the theta ($\theta$) angle distributions spanning $0^\circ$ to $180^\circ$, six bins for the phi ($\phi$) angle distributions spanning $-180^\circ$ to $180^\circ$, and ten bins for the alpha ($\alpha$) angle distributions ranging from 0$^\circ$ to 360$^\circ$. The invariant-mass binning was implemented in a two-step process. The acceptance calculation was performed using 7 coarse invariant-mass bins to ensure sufficient simulation statistics in the 7D phase space. Each of these bins was subsequently divided into two sub-bins, as described in Section \ref{sec:bin-centering}, yielding an effective binning of 14 invariant-mass bins used for the extraction of the cross sections.

\subsection{Electron Identification}
\label{sctn-EID}

A series of requirements was applied to multiple detector responses to identify negatively charged tracks as electron candidates. The cuts were designed to discriminate against minimum-ionizing particles, such as $\pi^-$s. The CLAS12 Event Builder (EB)~\cite{Ziegler:2020gsr} protocol first assigned electron identification to tracks with responses in the HTCC and ECAL that satisfied the criteria in Table~\ref{tab:eb_cuts}, with a geometrically matched associated hit in the FTOF based on DC tracking. After the EB identified electron candidates, additional cuts were applied to the data to select a refined sample for the data analysis.

\begin{table}[htp]
\begin{center}
\begin{tabular}{|c|c|} \hline
 Cut                       &  Limits \\ \hline
 Electric Charge           & negative \\ \hline
 Number of Photoelectrons  & $N_{\rm phe} > 2$ \\ \hline
 Min. PCAL Energy          & $E_{\rm dep} > 60$~MeV \\ \hline
 Sampling Fraction vs.~$p$ & $\pm5\sigma$ \\ \hline
\end{tabular}
\caption{The EB electron assignment requirements. The ECAL sampling fraction used by the EB was parameterized as a function of the total energy deposited in the ECAL. Note that the PCAL is the first module stack of the ECAL.}
\label{tab:eb_cuts}
\end{center}
\end{table}

Electrons in CLAS12 can only be identified in the FD given the coverage of the ECAL. The HTCC aids in reducing $\pi^-$ contamination in the electron sample for candidate tracks up to $\approx 4.9$~GeV (the threshold for $\pi^-$ to begin generating an HTCC signal in the CO$_2$ radiator gas). Up to the $\pi^-$ momentum threshold it is sufficient to cut on the number of photoelectrons ($N_{\rm phe}$) produced in the detector. A good electron candidate track will typically produce more than two photoelectrons (the CLAS12 trigger threshold), which was the minimum threshold for this cut. This cut was automatically enforced when using the EB to select electrons. 

\begin{figure}[ht]
  \centering
  \includegraphics[width=0.8\columnwidth]{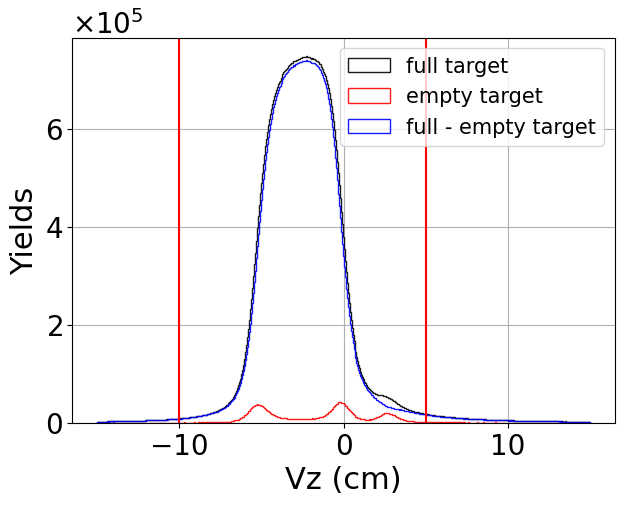}
  \caption{The $z$ component of the vertex position showing the full and normalized empty target distributions for all six sectors of the CLAS12 FD. The red histogram is from empty target runs, the black histogram is from the full target runs, and the blue histogram shows the difference. The vertical lines represent the cut limits.}
  \label{fig:vz_empty_target_sub}
\end{figure}

Due to the $W$ range of interest in this analysis, particles with momentum below 1.5~GeV were excluded from electron candidate selection. Additionally, the trace-back of the electron track from the ECAL through the DC toward the target must point back to the location of the target cell. A cut on the vertex coordinate along the beamline was applied in the range from $-10$ to $+5$~cm to safely account for the trace-back resolution. Figure~\ref{fig:vz_empty_target_sub} shows this reconstructed vertex for both full and empty target runs. The empty target distribution was normalized by the ratio of the total full over empty target accumulated charges measured by the Faraday cup. The empty target distribution histogram has three distinct peaks. The first two from the left arise from contributions from the target cell entrance and exit windows. The rightmost peak is from an insulation layer that surrounds the target cell.

\begin{figure}[ht]
  \centering
  \includegraphics[width=0.95\columnwidth]{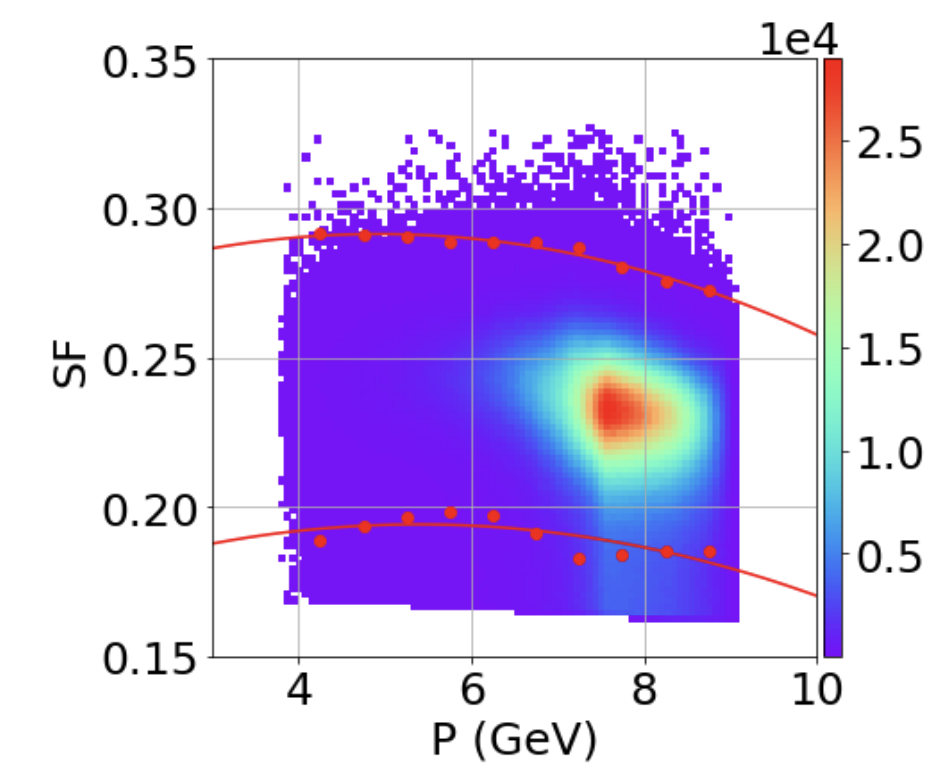}
  \caption{Sampling fraction vs. momentum for electrons for a representative sector using measured data. The red lines are from a second-order polynomial fit, which represents the applied $\pm$3.5$\sigma$ cut.}
  \label{fig:electron_sf_exp}
\end{figure}

The ECAL sampling fraction (SF) is a measure of the incident electron energy deposited in the active layers of the calorimeter. For the ECAL the SF has a value of about 25\% with a weak dependence on the electron energy. It is defined as the ratio of the measured electron energy in the calorimeter to that determined from the track reconstruction. The SF cuts were developed separately for the six sectors of the FD based on a fit of the SF vs. momentum. The nominal cuts were defined cutting $\pm 3.5 \sigma$ about the mean. This cut was tighter than the nominal EB cut to further refine out electron selection. A similar procedure was applied to the MC. Figure~\ref{fig:electron_sf_exp} shows the SF vs. momentum distribution for a representative sector along with the cuts.

\begin{figure}[ht]
  \centering
  \includegraphics[width=0.95\columnwidth]{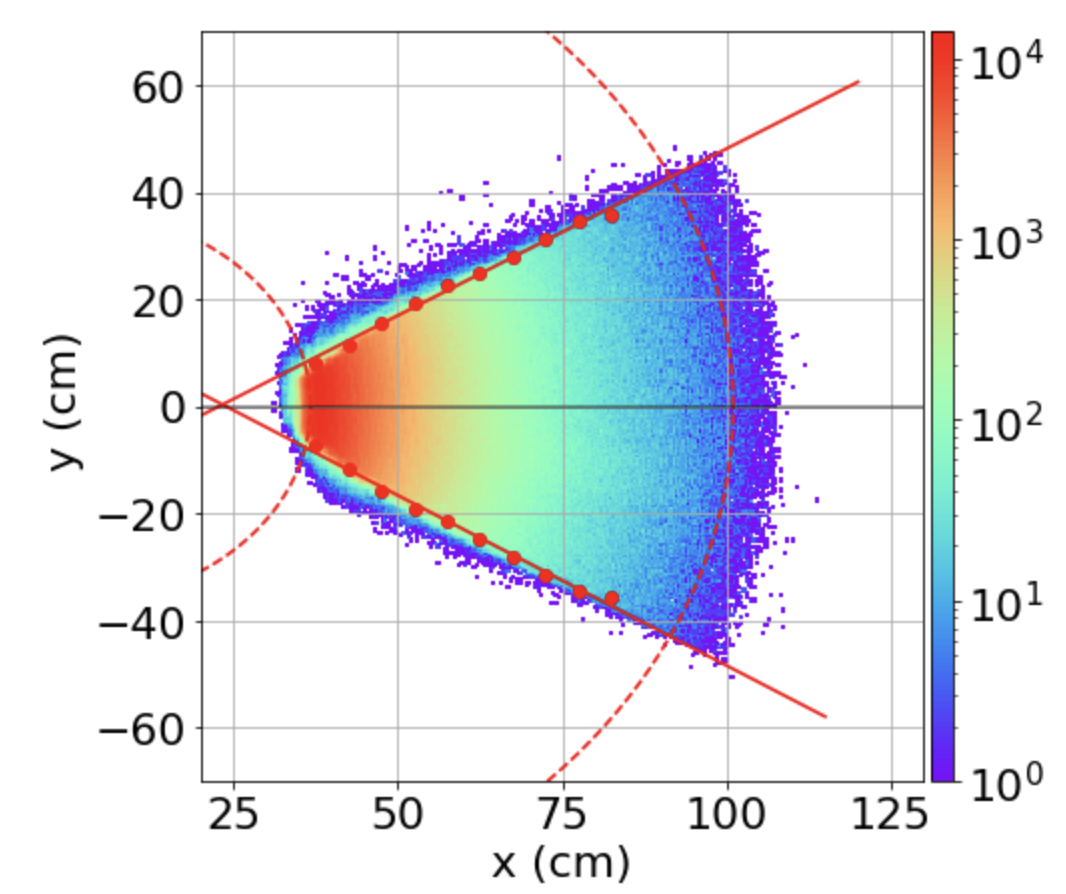}
  \caption{Measured $y$ vs. $x$ hit coordinates for electron candidates for a representative FD sector projected onto a plane in the innermost DC detector with the defined fiducial cuts overlaid.}
  \label{fig:dc_fid_cuts}
\end{figure}

The last aspect of defining the sample of scattered electrons for the two-pion dataset was to select a matched fiducial volume for both measured and simulated data. The CLAS12 detector includes physical gaps, such as the forward-angle hole resulting from beamline components, sector gaps due to the torus coils, and malfunctioning detector elements. Additionally, factors such as coil re-scattering and magnetic field distortions can affect particles near the edges of the detector. When an electromagnetic shower created by an electron occurs near these edges, energy leakage can lead to incomplete energy reconstruction and a misidentification of electrons. To mitigate these effects, a cut was implemented to remove the edge regions of the ECAL using information from the reconstructed ECAL cluster coordinate. In addition, the outer edges of the DC near the torus coils were cut out to ensure high and uniform recording of hits. The cuts for a representative DC layer are shown in Fig.~\ref{fig:dc_fid_cuts}. Finally, at this stage, coordinate cuts on the FTOF and ECAL detectors were implemented to remove malfunctioning detector elements. The fiducial volume cuts were matched in both measured and simulated data.

\subsection{Charged Hadron Identification}

In the two-pion electroproduction channel, there are three hadrons in the final state. In the missing $\pi^-$ topology (used in this analysis), only reconstructed protons and $\pi^+$s were directly considered and the four momentum of the $\pi^-$ was reconstructed using the known four momenta of all other particles. The hadrons detected by both the FD and the CD were used in this analysis. The initial identification of charged hadrons was based on the CLAS12 EB~\cite{Ziegler:2020gsr} that defined track candidates by matching a track to a hit in either the FTOF or CTOF. The particle identification (PID) assignment was based on time-of-flight. For each track of momentum $p$, the measured flight time from the event vertex to the TOF system, $TOF_\text{m}$, was compared to the calculated time, $TOF_\text{c}$, for a pion, kaon, or proton of identical momentum. Cuts were placed on the difference between the measured and calculated times, $\Delta t = TOF_\text{m} - TOF_\text{c}$. The particle species assumption that minimized $\Delta t$ was assigned as the particle type.

\begin{figure}[ht]
  \centering
  \includegraphics[width=1.0\columnwidth]{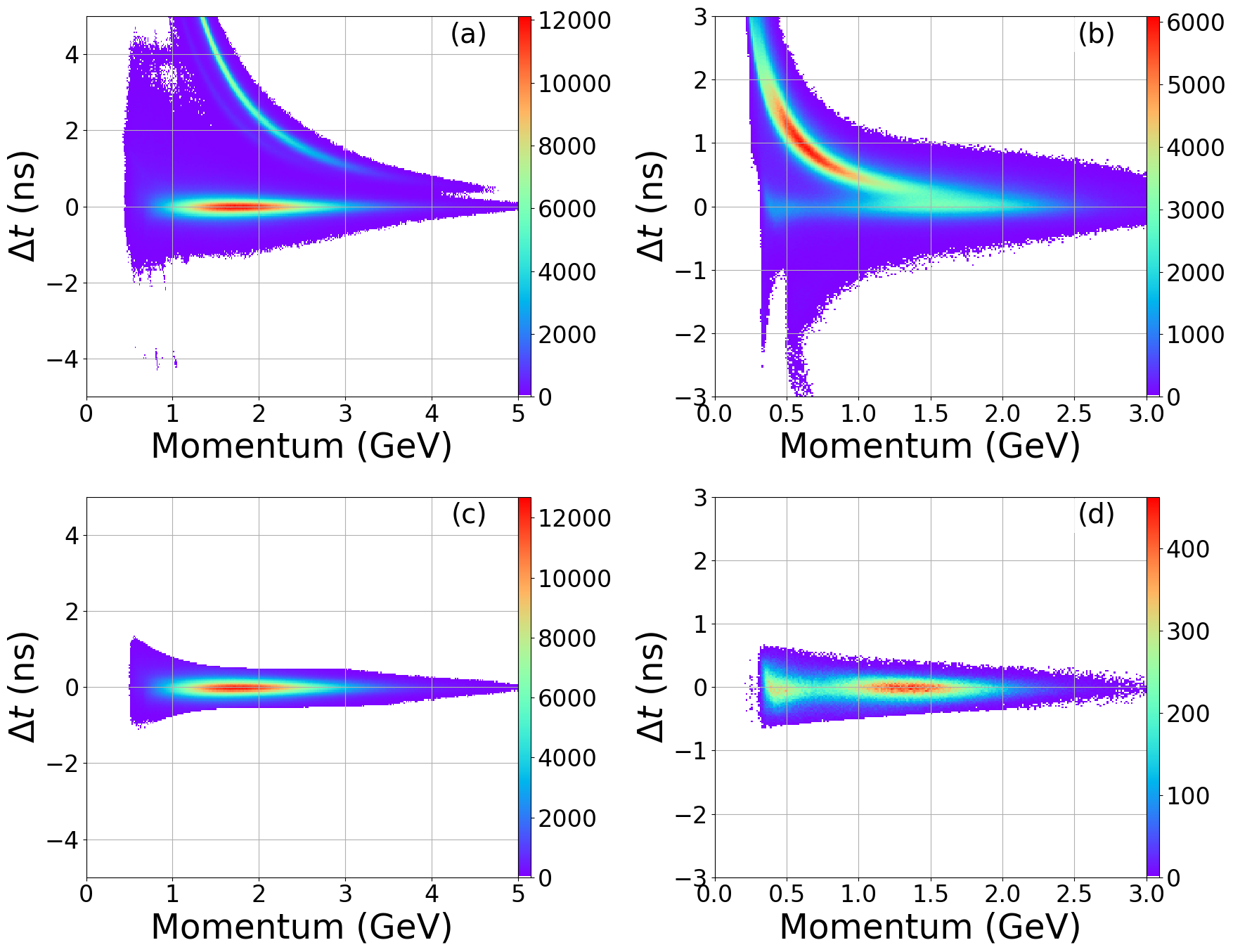}
  \caption{$\Delta t$ vs. momentum distributions for positively charged particles assuming the mass of a proton before (top) and after (bottom) PID cuts, using measured data. The FD hadrons are shown in the left plots and the CD hadrons in the right plots.}
  \label{fig:dt-protons}
\end{figure}

Charged particles with $p < 0.2$~GeV in the CD and 0.5~GeV in the FD were rejected as they had very small acceptance in CLAS12 due to the magnetic field configuration. The difference in the $z$ component of the vertex position between each of the hadron candidates and the electron vertex was required to be within $\pm 20$~cm for both FD and CD hadrons. The same loose clean-up cuts were applied for both the data and MC to remove scattering sources along the beamline not associated with the liquid-hydrogen target. It should be noted that the trace-back resolution of CLAS12 is at the level of 0.5--1~cm in the FD and 1--2~mm in the CD. Finally, $\Delta t$ cuts were applied based on the difference between the measured and calculated flight times using either the FTOF or CTOF to remove outliers from the hadron identification sample. Figure~\ref{fig:dt-protons} shows $\Delta t$ vs. momentum distributions for all positively charged tracks assuming a proton mass for both FD and CD hadrons before and after the full set of hadron PID cuts. Separate parameters were defined for the FD and CD $\Delta t$ cuts for protons and $\pi^+$s. 

\begin{figure}[ht]
  \centering
  \includegraphics[width=0.95\columnwidth]{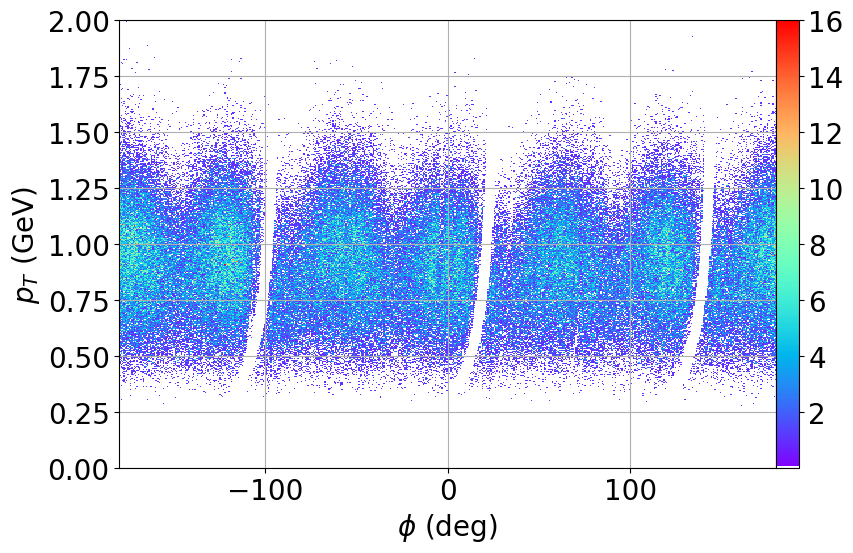}
  \caption{Transverse momentum $p_T$ vs. $\phi$ for CD protons (after PID and two-pion selection cuts) from data. The three azimuthal gaps correspond to the borders between the internal modules within the CVT tracking system.}
  \label{fig:mom_vs_phi_prot}
\end{figure}

As was done for the electrons, DC fiducial cuts were defined for the FD charged hadrons to remove tracks near the edges of the tracking detectors. These edge regions suffer from reduced hit resolution and efficiency. For charged hadrons in the CD, geometric cuts were placed around the internal azimuthal gaps between the CVT modules using the track transverse momentum ($p_T = \sqrt{p_x^2 + p_y^2}$) and azimuthal angle ($\phi$). Figure~\ref{fig:mom_vs_phi_prot} shows the transverse momentum $p_T$ vs. $\phi$ distribution for the CD protons in the final two-pion sample. These hadron fiducial cuts were applied to both measured and MC data.

\subsection{Energy Loss and Momentum Corrections}

The measured charged particle momenta in CLAS12 have inaccuracies due to potential geometrical misalignments of the tracking detectors relative to the magnetic fields, systematic reconstruction and calibration biases, and inaccuracies in the magnetic field maps for the torus and solenoid used for the charged particle tracking. To reduce the associated systematics, the measured momenta from CLAS12 were corrected for the FD and CD hadron energy loss through the passive detector materials using simulation studies and momentum corrections were applied for the different final state particles based on exclusive event reconstruction of multiple final states (e.g. elastic $ep$, $\pi N$, and $\pi^+\pi^-p$).  In the FD of CLAS12, the momentum resolution is $\Delta p/p \sim 0.5 - 1\%$. For the CD, the momentum resolution is slightly worse at 5\% due to the short path length to the detectors and the tracking resolution. After the energy loss and momentum corrections, the residual distortions of the $\pi^-$ missing mass squared (MMSQ) distribution were at a level below $\pm 5$~MeV over the full kinematic phase space of the data. The systematic uncertainty for this correction is detailed in Section~\ref{sec:syst}.

\subsection{Event Selection} 
\label{sec:event-sel}

This analysis uses only the missing $\pi^-$ topology to select the two-pion events using cuts on the $\pi^-$ MMSQ distribution. The use of the missing $\pi^-$ topology is advantageous as it has higher statistics than the other topologies, because in the inbending CLAS12 detector setup, $\pi^-$s in the FD are harder to detect as negatively charged particles bend toward the beamline due to the magnetic field. Consequently, a significant number of these particles do not reach the FTOF, which is a requirement for charged particle identification in the EB.

\begin{figure}[ht]
  \centering
  \includegraphics[width=0.85\columnwidth]{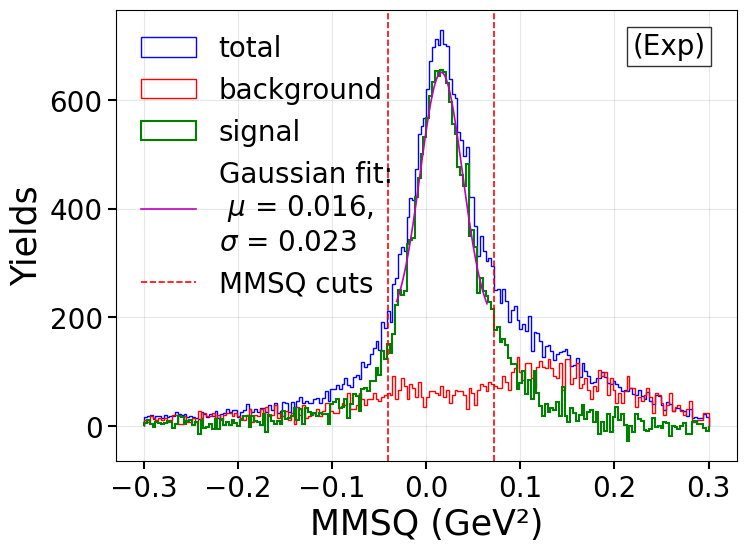}
  \\[1ex]
  \includegraphics[width=0.85\columnwidth]{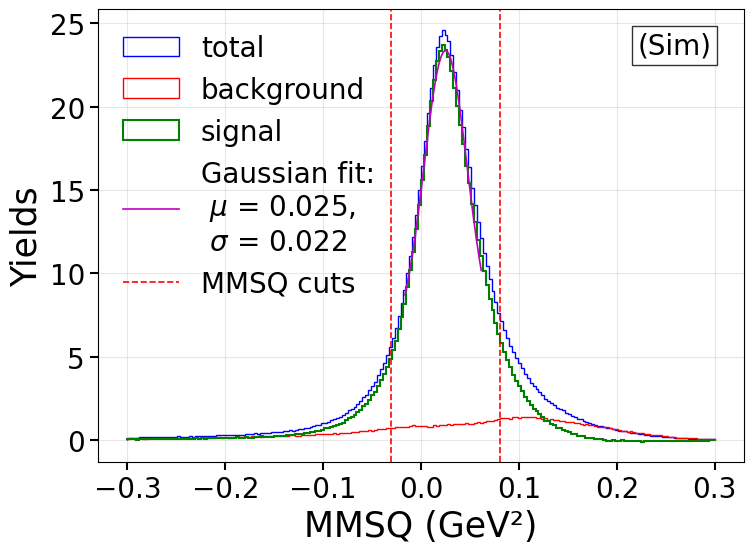}
  \caption{MMSQ $\pi^{-}$ distributions for the representative bin $W \in [1.70, 1.75]$~GeV and $Q^2 \in [3.5,4.2]$~GeV$^2$ - experiment (top) and MC (bottom).}
  \label{fig:mmsq_cuts}
\end{figure}

The MMSQ cuts are $(W,Q^2)$ dependent. For each kinematic bin, the MMSQ distribution was first corrected by subtracting a background contribution determined from the fully exclusive topology by applying cuts on the three MMSQ variables: missing proton, missing $\pi^+$, and fully exclusive. The resulting background sample was then scaled to the $\pi^-$ MMSQ distribution by matching the integral of the tails on either side of the distributions, as illustrated in Fig.~\ref{fig:mmsq_cuts}, which shows an example for the kinematic bin $W \in [1.70,1.75]$~GeV and $Q^2 \in [3.5,4.2]$~GeV$^2$. The same procedure was applied for measured and MC data.

\subsection{Corrected Event Yields}
\label{cha:corr}

\subsubsection{Reconstruction Efficiency Corrections} 
\label{sec:part_eff}

Ideally, the particle reconstruction efficiency of the experiment should be the same as for the MC simulation. However, for this analysis, a full matching of the simulation to the experiment was not achieved, which would result in unmatched measured and MC particle reconstruction efficiencies if not taken into account. 

A data-driven method was developed and used to compare particle reconstruction efficiencies between experiment and MC. Where differences were observed, efficiency correction factors were applied on an event-by-event basis, with these factors defined as
\begin{equation}
    {\cal E} = {\cal E}(p,\theta, \phi)_{\rm exp} = \frac{\varepsilon(p,\theta, \phi)_{\rm exp}}{\varepsilon(p,\theta, \phi)_{\rm MC}}.
\label{eq:eff_corr_formula}
\end{equation}
Here $\varepsilon_{\rm MC}$ is the product of the individual hadron particle reconstruction efficiencies in MC, which are defined as the ratio of the number of particles reconstructed in the exclusive topology $ep \to e'\pi^+\pi^-p'$ to the number reconstructed in the missing topology. Similarly, $\varepsilon_{\rm exp}$ is the corresponding ratio based on the experimental reconstruction. These efficiency factors were calculated by mapping the momentum $p$, polar angle $\theta$, and azimuthal angle $\phi$ dependence of each measured hadron in the laboratory frame. The number of particles in these kinematical distributions, for both experiment and MC, was determined from the corresponding integrals of the background-subtracted MMSQ distributions within $\pm 2\sigma$ of the mean using a Gaussian fit around the MMSQ peak. To minimize the systematic uncertainty of this process, background subtractions were first carried out in each $W$ and $Q^2$ bin (see Fig. \ref{fig:mmsq_cuts}), as the background and signal shapes are both $W$ and $Q^2$ dependent, and then recombined for the full hadron $(p,\theta, \phi)$ mapping.

\begin{figure}[!ht]
  \centering
 \includegraphics[width=0.95\columnwidth]{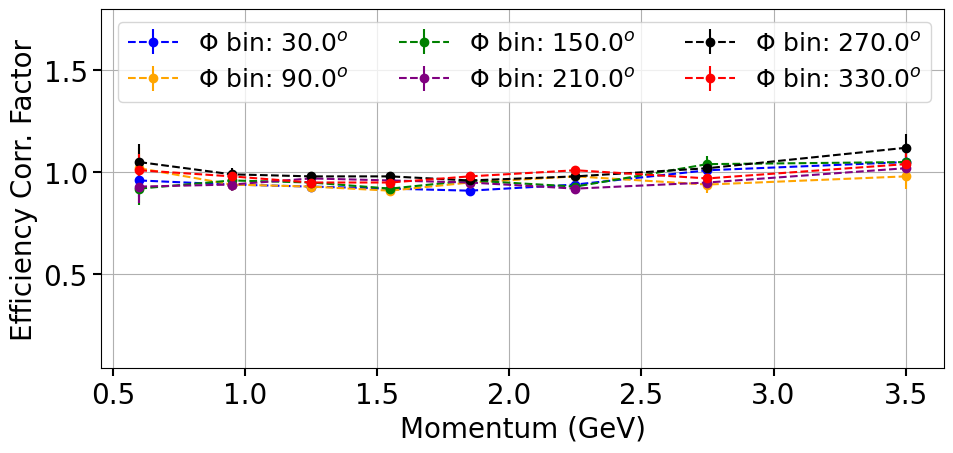}
 \includegraphics[width=0.95\columnwidth]{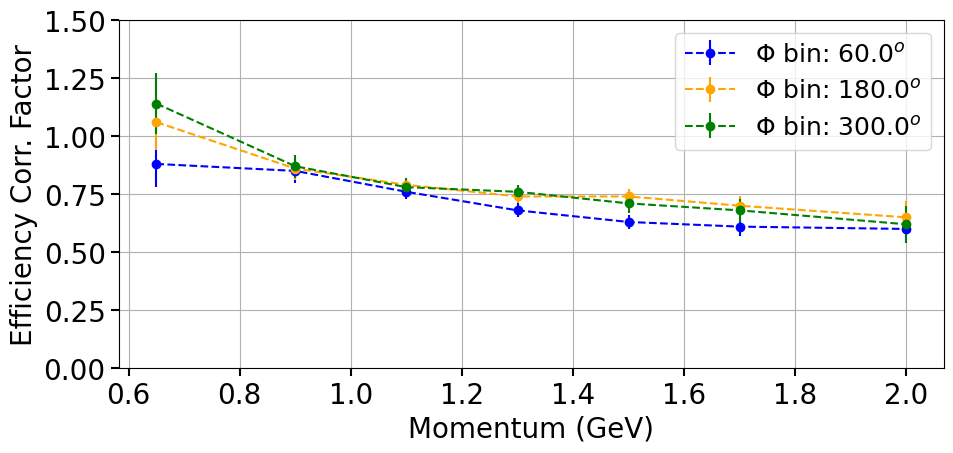}
  \caption{Momentum dependence of the proton efficiency correction factors for $\theta < 37^{\circ}$ (top) and $\theta \geq 37^{\circ}$ (bottom).}
  \label{fig:eff_fact_prot}
\end{figure}

\begin{figure}[!ht]
  \centering
 \includegraphics[width=0.95\columnwidth]{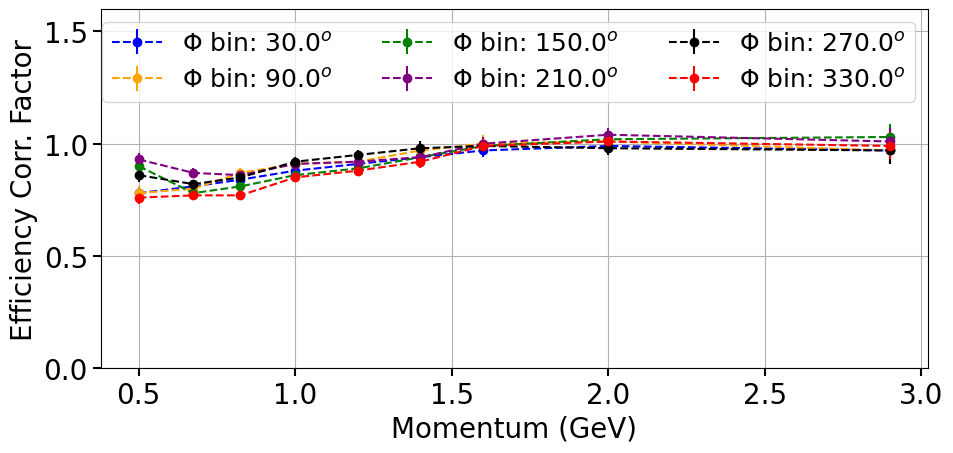}
 \includegraphics[width=0.95\columnwidth]{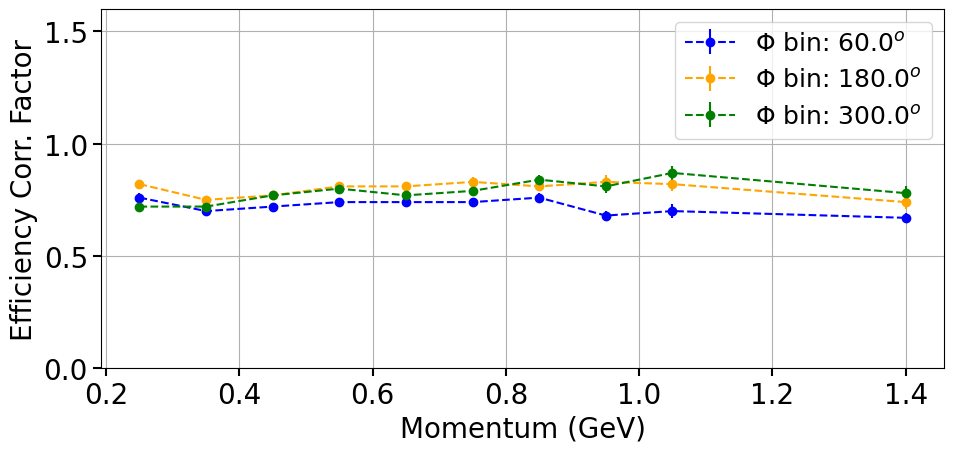}
  \caption{Momentum dependence of the $\pi^+$ efficiency correction factors for $\theta < 37^{\circ}$ (top) and $\theta \geq 37^{\circ}$ (bottom).}
  \label{fig:eff_fact_pip}
\end{figure}

Figures~\ref{fig:eff_fact_prot} and \ref{fig:eff_fact_pip} show the momentum and $\phi$ dependence of the individual hadron efficiency correction factors for all protons and $\pi^+$s, respectively, in the integrated polar angle $\theta < 37^{\circ} $ and $\theta \geq 37^{\circ}$ ranges. The correction factor for protons with $\theta < 37^{\circ}$ is close to unity, indicating that the reconstruction efficiency is similar in both experiment and MC, however, for protons with $\theta \geq 37^{\circ}$ the correction factors for all three $\phi$ bins are similar but all show a linear decrease with momentum, reaching $\sim 0.6$ at 2~GeV. For positive pions with $\theta < 37^{\circ}$, the efficiency correction factors are close to unity for $p \gtrsim 1.5$~GeV, but fall to around 0.8-0.9 for momenta approaching 0.5~GeV. For $\theta \geq 37^{\circ}$, the $\pi^+$ efficiency correction factors are at a value of about 0.7 and relatively flat vs. momentum. The systematic uncertainty associated with this correction to the data yields is detailed in Section~\ref{sec:syst}.

\subsubsection{Background Subtraction}
\label{sec:background}

The two-pion electroproduction data, even after the application of all particle-identification and event-selection criteria, still contained residual background within the MMSQ cuts. This background originates from the three-pion channel and is apparent in the $\pi^-$ ${\rm MMSQ} > 0$ tail region. In the three-pion channel, 
\begin{equation*}
ep \to e'\pi^+\pi^-\pi^0p',
\end{equation*}
the undetected neutral pion distorts the reconstructed MMSQ distribution.

\begin{figure}[!ht]
  \centering
 \includegraphics[width=0.95\columnwidth]{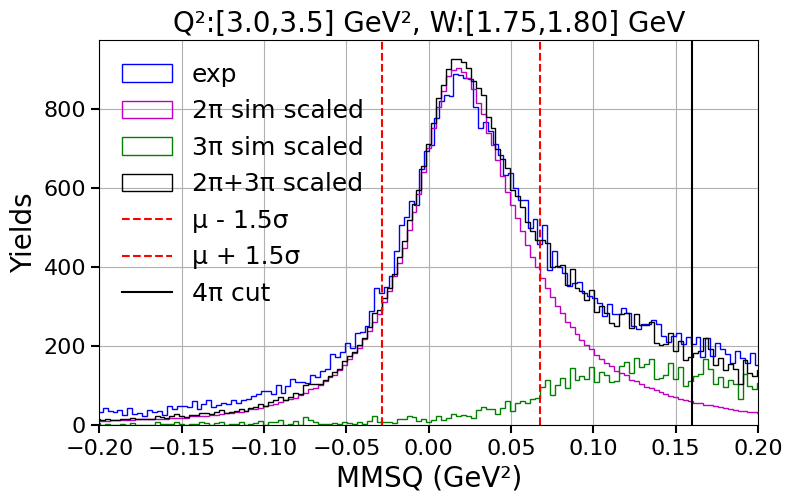}
 \\[1ex]
 \includegraphics[width=0.95\columnwidth]{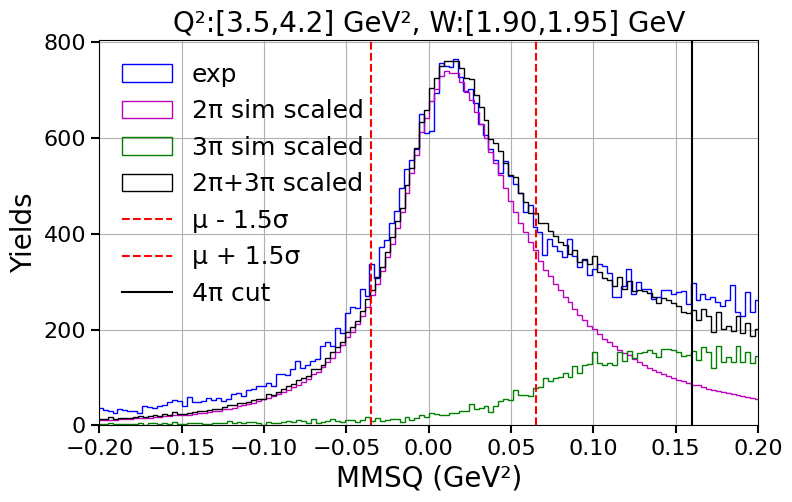}
  \caption{MMSQ distributions for the $\pi^-$ for representative $(W,Q^2)$ bins showing the experimental data (blue), the scaled two-pion and three-pion simulations (magenta and green), and their sum (black). The dashed red lines indicate the $\pm 1.5\sigma$ region about the mean ($\mu$), and the solid black line denotes the four-pion threshold cut applied at 0.16~GeV$^2$.}
  \label{fig:mmsq_background_w_q2_bins}
\end{figure}

To account for this contribution, three-pion events were simulated over the full kinematic range of this analysis ($2.4 ~\text{GeV}^2\le Q^2 \le 8.0~\text{GeV}^2$, $1.4 ~\text{GeV}\le W \le 2.1~\text{GeV}$) using a phase space generator merged with 45~nA background files (see Section~\ref{sec:mc} for details on background merging). For each $(W,Q^2)$ bin, the three-pion MC contribution was normalized to the measured data on the right-hand side of the $\pi^-$ MMSQ distribution from $\mu$\,+\,1.5\,$\sigma$ to the four-pion threshold (0.16~GeV$^2$), so that the sum of the scaled two-pion and three-pion MC distributions reproduced the measured data. Figure~\ref{fig:mmsq_background_w_q2_bins} shows representative MMSQ distributions for two different $(W,Q^2)$ bins with the two-pion and three-pion MC spectra accordingly normalized to the experimental data.

\subsection{Cross Section Evaluation}

The expression for the seven-fold differential electron scattering cross section specific to the charged two-pion electroproduction channel is given by  

\begin{multline}
   \frac{d^7\sigma}{dWdQ^2d^5\tau} = 
   \frac{1}{A ~R ~{\cal L}} \frac{\left(\frac{N^{\cal E}_{\rm full}}{Q_{\rm full}} - \frac{N^{\cal E}_{\rm empty}}{Q_{\rm empty}}\right)}{\Delta W \Delta Q^2 \Delta^5\tau} BC_{\text{corr}},
   \label{equ:cs-formula}
\end{multline}
\noindent
where $N^{\cal E}_{\rm full}$ and $N^{\cal E}_{\rm empty}$ are the numbers of event-by-event efficiency corrected (see Eq.~(\ref{eq:eff_corr_formula})) two-pion events selected inside the 7D bin $\Delta W \!\Delta Q^2\! \Delta^5 \tau$ for the filled and empty liquid-hydrogen target, respectively. Similarly, $Q_\text{full}$ and $Q_\text{empty}$ are the total charges collected in the Faraday cup for the full and empty target runs, respectively. For this analysis, those charges were 28.78~mC for the full target runs and 2.23~mC for the empty target runs. $A$ is the acceptance factor obtained from the MC simulations,
$R$ is the radiative correction factor, and $BC_{\text{corr}}$ is the bin-centering correction factor. $\Delta W$, $\Delta Q^2$, and $\Delta^5 \tau$ are the bin sizes of the analysis variables out of which $\Delta W$ and $\Delta Q^2$ are given by the electron scattering kinematics and $\Delta^5 \tau$ by the hadronic 5D phase space cell, 
\begin{equation}
    \Delta^5 \tau=\!\Delta M_{h_{1}h_{2}}\Delta M_{h_{2}h_{3}}\Delta \Omega_{h_{1}}\!\Delta \alpha_{h_{1}}.
\end{equation}
${\cal L}$ is the beam-target luminosity per unit accumulated beam charge given by
\begin{equation}
    {\cal L} = \frac{l\rho N_\text{A}}{q_\text{e} M_\text{H}},
\end{equation}
\noindent
where $q_\text{e}$ is the electron charge, $l$ length of the target (5 cm), $\rho$ is the density of liquid hydrogen (0.0708~g/cm$^3$), $M_\text{H}$ is the molar mass of hydrogen ($1.0079$~g/mol), and $N_\text{A}$ is Avogadro's constant. For our dataset, $\mathcal{L} \approx 1.32\times10^{42}~\text{cm}^{-2}\text{C}^{-1}$ and $   \mathcal{L}_\text{int} = \mathcal{L}\times Q_\text{full} \approx 3.80\times10^{40}~\text{cm}^{-2} \approx 38.0~\text{fb}^{-1}$. For the empty target runs, the liquid-hydrogen is removed, but cold hydrogen gas remains in the target cell. The density of the gas is $\rho_\text{gas}$ = 0.0010~g/cm$^3$. Therefore, the adjusted density $\rho_\text{adj} = \rho - \rho_\text{gas}$ = 0.0697~g/cm$^3$ was used for the final calculations.

\section{Detector Simulations}
\label{sec:mc}

In this analysis, the MC simulation was performed using the TWOPEG event generator developed for charged two-pion electroproduction~\cite{2017skorodumina}. This realistic description of the production cross sections has been used in previous CLAS two-pion electroproduction cross section analyses \cite{Fedotov:2008aa,CLAS:2017fja,CLAS:2018fon,CLAS:2023mfc,Trivedi:2018rgo}.

Events generated with TWOPEG were passed through a standard multi-stage procedure of detector simulation and event reconstruction. The same software was used for both experiment and MC event reconstruction as well as the same analysis code, with all analysis cut parameters appropriately matched between experiment and MC. The CLAS12 detector response was modeled using Geant4-based~\cite{GEANT4:2002zbu} simulation software called GEMC~\cite{Ungaro:2020xlc}, which includes the full CLAS12 geometry, magnetic fields, tracking, and detector inefficiencies. 

TWOPEG employs a weighted event generation approach to throw events uniformly in the selected kinematic range. The simulation sample in this analysis covered a slightly larger $W$ [1.35:2.15]~GeV and $Q^2$ [1.95:9.0]~GeV$^2$ range than the experiment. The cross section values were assigned as weights to each generated event. These cross sections were derived from the phenomenological JM model~\cite{Mokeev:2015lda}, which was constrained by fitting the measured cross sections from the CLAS detector. In kinematic regions where experimental cross sections were not available, the weights were estimated by interpolating and extrapolating known structure functions. TWOPEG was designed for a wide kinematic range and is suitable for the CLAS12 energy range as well. Events can be generated, including radiative effects, on the basis of the Mo-Tsai approach~\cite{Mo:1968cg}.

\begin{figure}[!ht]
  \centering
 \includegraphics[width=0.9\columnwidth]{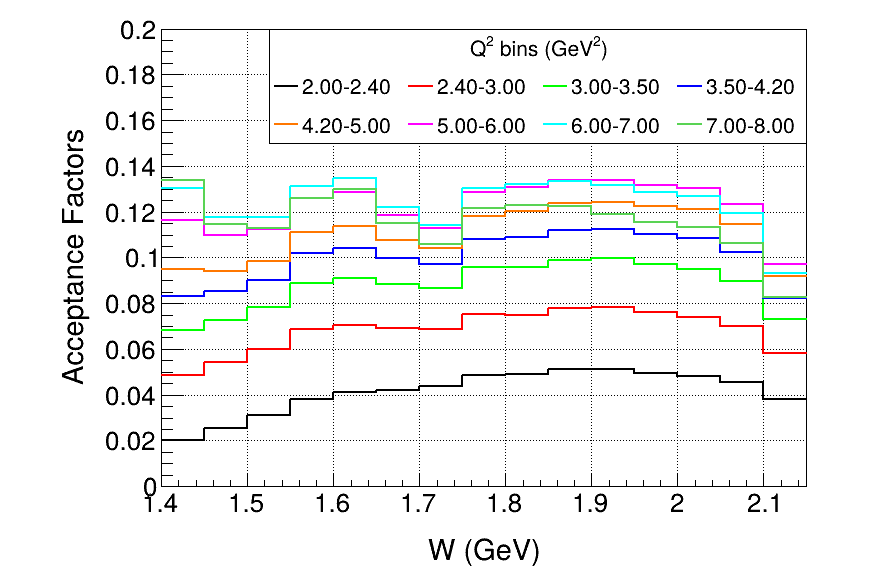}
 \includegraphics[width=0.9\columnwidth]{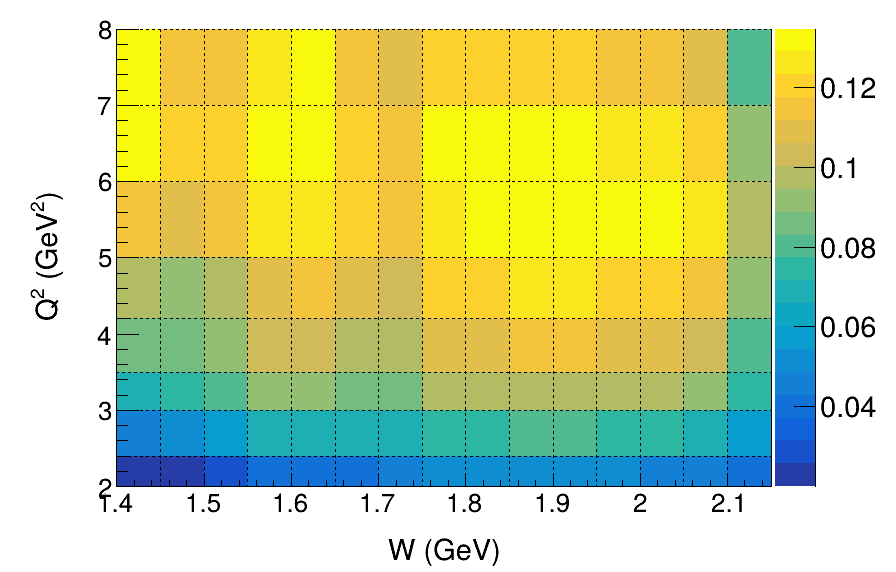}
  \caption{(Top) Acceptance factors for various $Q^2$ bins in the range $W \in [1.4, 2.15]$~GeV. (Bottom) Color-coded acceptance factors in bins of $Q^2$ vs. $W$.}
  \label{fig:acc_corr}
\end{figure}

To determine the two-pion cross sections, it is essential to account for events that were not measured by CLAS12 or excluded by the event selection or other analysis cuts. This was achieved using the detector acceptance factor $A$ incorporated into the cross section formula of Eq.~(\ref{equ:cs-formula}). This factor, that accounts for both geometrical acceptance and detector inefficiencies of any kind otherwise not accounted for, was defined as

\begin{equation}
    A(\Delta W, \Delta Q^2, \Delta^5 \tau) = \frac{N_{\rm rec}}{N_{\rm gen}}.
    \label{equ:acc_corr_fact}
\end{equation}
\noindent
Here, $N_{\rm gen}$ is the cross section weighted number of two-pion events generated within a given 7D kinematic bin, and $N_{\rm rec}$ corresponds to the cross section weighted number of two-pion events reconstructed in each bin. The weights are event-specific cross sections generated by TWOPEG at each kinematic point. The resulting acceptance factor distributions as a function of $W$ and $Q^2$ are shown in Fig.~\ref{fig:acc_corr}.

For this analysis, nearly $10^{10}$ events were generated across the full kinematic phase space. The factor $A$ implicitly accounts for bin migration effects where events were reconstructed in bins different from those in which they were generated.

\subsection{Background Merging in Monte Carlo} 
\label{sec:bkg_merge}

Due to the hit occupancies in the tracking detectors (up to 10\% in the DC and up to 2\% in the CVT), the RG-A data showed non-negligible inefficiencies in the charged particle reconstruction. These occupancies are associated with ``background noise" hits in the detector in addition to the signals from the passing charged particles. These extra hits affect the track reconstruction due to confusion of which hits belong to the tracks. This leads to a reduction of the track reconstruction efficiency that depends on the luminosity. The event-by-event simulation for this analysis was on the other hand luminosity independent and, hence, employed a background merging option in order to best match the backgrounds in the various CLAS12 detector subsystems~\cite{clas12_bkgm}. 

Background hit merging for the DC, FTOF, ECAL, CTOF, and CVT systems was developed based on RG-A data collected with a random trigger at the production beam currents. These data events were merged with the MC event-by-event. As the background was determined based on actual data from CLAS12 taken in the exact same conditions as the data used for analysis, it matches well the overall background distribution in the detector to model the tracking inefficiency across the full FD and CD acceptance as a function of luminosity. This merging of background into MC events was performed at the level of raw ADC and TDC hits. The systematic uncertainty assigned to the track reconstruction efficiency is discussed in Section~\ref{sec:syst}.

\subsection{TWOPEG Cross Section Adjustments}
\label{sec:cs_scaling_factor}

TWOPEG was originally developed using the available cross sections extracted from CLAS data for $Q^2 \le 1.3$~GeV$^2$. For higher-$Q^2$ CLAS and the new CLAS12 RG-A datasets, the $Q^2$ dependence of the extrapolated TWOPEG cross sections needed adjustment. The original $Q^2$ evolution in TWOPEG was based on

\begin{equation}
    \frac{d^5\sigma}{d^5\tau} = \frac{d^5\sigma}{d^5\tau}\left(1.3\;\text{GeV}^2\right) \frac{F_0(Q^2)}{F_0(1.3\;\text{GeV}^2)},
\end{equation}
where $d^5\sigma/d^5\tau$ is the five-fold differential hadronic cross section and
\begin{equation}
   F_0 =  \frac{1}{\left(1+\frac{Q^2}{0.7\;\text{GeV}^2}\right)^2}
   \label{equ:F_app}
\end{equation}
is the $Q^2$-dependent scaling function. 

The ratio of higher $Q^2$ CLAS data to the TWOPEG cross sections was fitted by 
\begin{equation}
   F_\text{new} =  \frac{\left(F_0(Q^2)\right)^{a}}{\left(F_0(1.3\;\text{GeV}^2)\right)^{b}},
   \label{equ:F_app1}
\end{equation}
\noindent
where $a$ and $b$ are the parameters determined by the fit, and the cross sections measured from this analysis were used to extend the fit up to 8~GeV$^2$.

\begin{figure*}[!ht]
  \centering
  \includegraphics[width=0.80\textwidth]{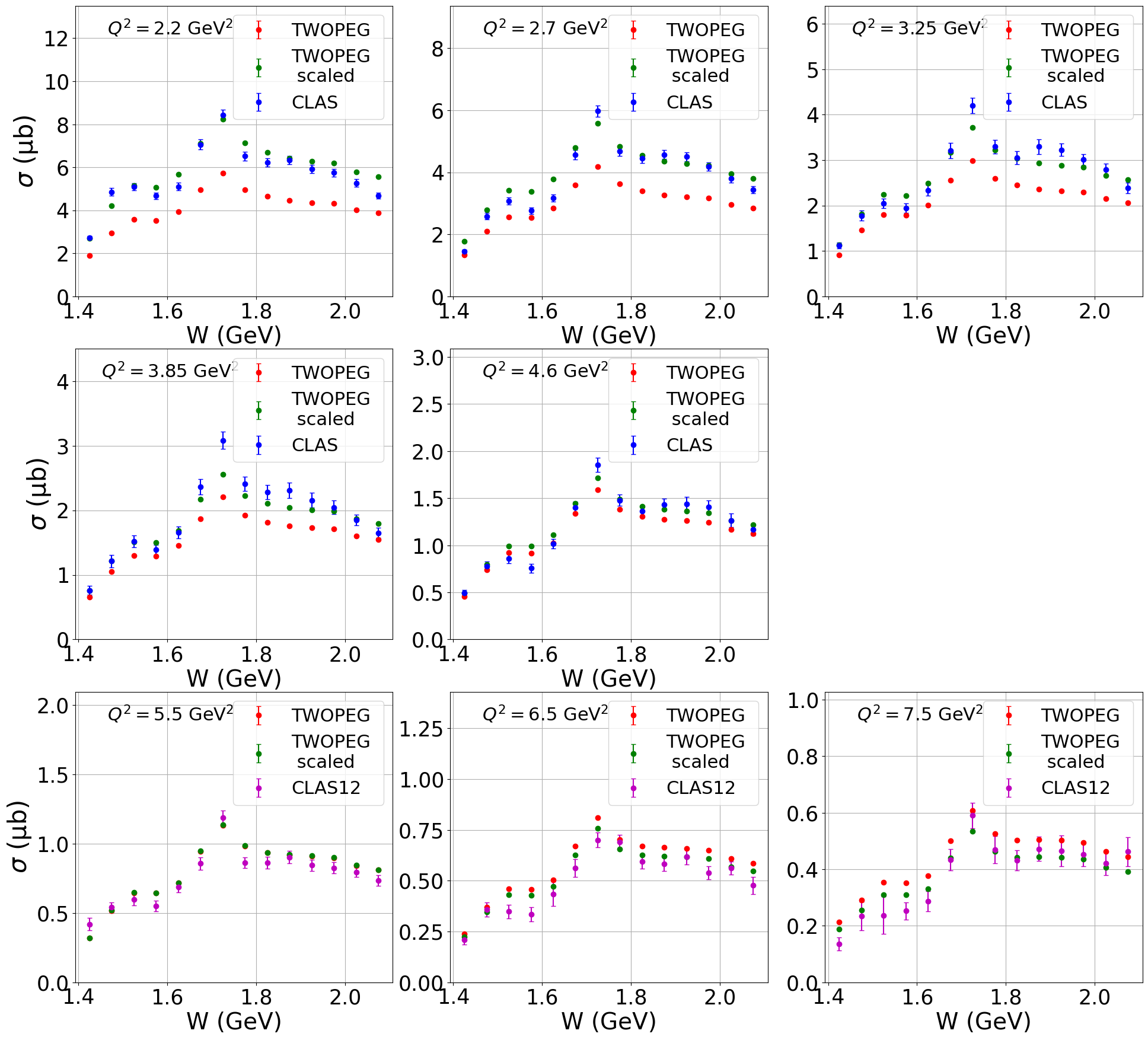}
  \caption{Integrated cross sections for various $Q^2$ bins comparing the measured CLAS/CLAS12 data (blue/magenta points), the original TWOPEG model (red points), and the $Q^2$-dependent scaled TWOPEG (green points) cross sections.}
  \label{fig:scaling_cs}
\end{figure*}

Figure~\ref{fig:scaling_cs} compares the original TWOPEG generator cross sections, the measured cross sections from CLAS~\cite{Trivedi:2018rgo}, CLAS12, and the CLAS12-adjusted TWOPEG cross sections for all $Q^2$ bins. This adjustment is particularly important for filling zero-acceptance bins (see Section~\ref{sec:hole-filling}), bin-centering corrections (see Section~\ref{sec:bin-centering}), and radiative corrections (see Section~\ref{sec:rad-corr}), all of which are modified using this new $Q^2$-dependent scaling function. Despite the improvement achieved by refitting the $Q^2$-dependent scaling, this effective one-fits-all approach does not distinguish between different $Q^2$ dependencies of the various background and resonance amplitudes. Hence some discrepancies remain between the scaled TWOPEG cross sections and the measured CLAS/CLAS12 cross sections. While the overall agreement is good, some bins show noticeable deviations, particularly in the regions associated with the resonance peaks. To account for these differences, a second adjustment step was applied in which the TWOPEG cross sections were further corrected using normalization factors for each $(W,Q^2)$ bin. The adjusted integrated TWOPEG cross section for each $(W,Q^2)$ bin was compared to the corresponding CLAS/CLAS12 measured cross section, and a multiplicative correction factor was derived from their ratio. This factor was then applied uniformly to all nine single-differential cross sections generated by TWOPEG within that $(W,Q^2)$ bin. This procedure ensured that the integrated TWOPEG cross sections match well to the data in every bin. This bin-by-bin scaling is particularly important to reduce the model-dependent hole-filling uncertainties, since kinematic regions with zero acceptance are populated using TWOPEG-generated events. 

\subsection{Monte Carlo Smearing}

In this analysis, the TWOPEG event generator was used to simulate exclusive charged two-pion electroproduction off the proton under the same conditions and selection criteria as in the experiment. Ideally, the resolution of the MMSQ peaks of the MC should match that of the experimental data. However, the CLAS12 simulation has not yet fully replicated the FD and CD responses. To ensure optimized yield extractions, efficiency studies, background analyses, and precise cross section extractions, it was necessary to smear the simulated MMSQ distributions to match the measured resolutions. 

\begin{figure}[!ht]
  \centering
 \includegraphics[width=0.9\columnwidth]{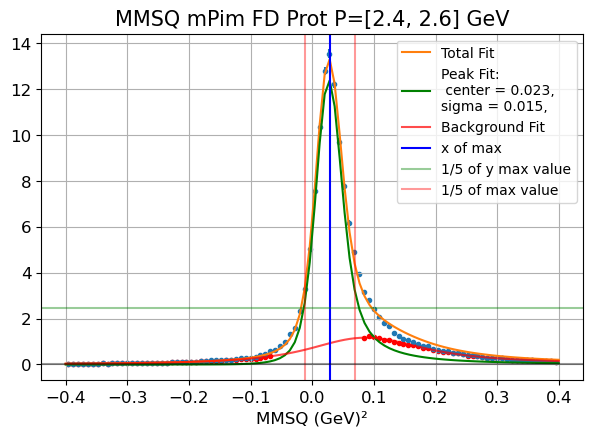}
 \includegraphics[width=0.9\columnwidth]{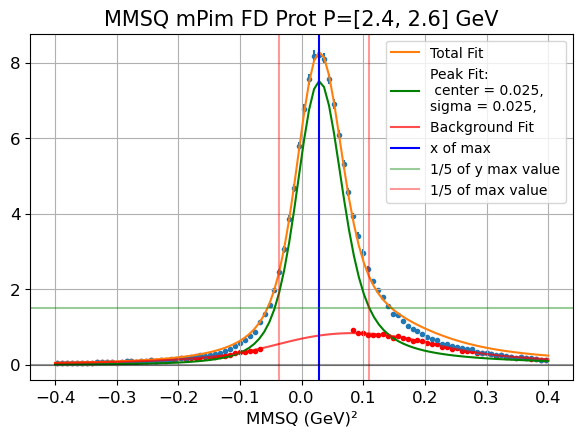}
 \includegraphics[width=0.9\columnwidth]{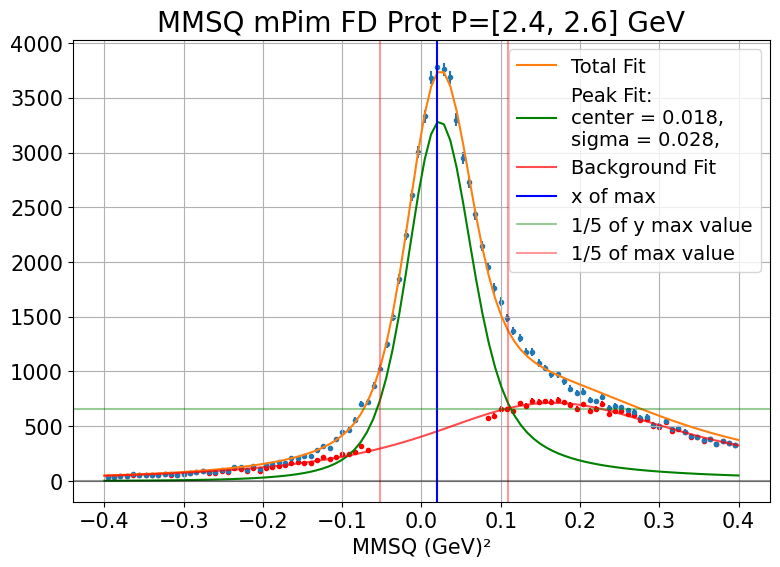}
  \caption{Typical $\pi^-$ MMSQ distributions for a representative FD proton momentum bin [2.4:2.6]~GeV from MC before smearing (top) and after smearing (middle). The bottom plot shows the measured distribution used to adjust the smearing parameters. In all plots, the blue points represent the signal and the red points the background, whereas the red curves represent the background, the green curves the signal, and the orange curves the total fits. The red vertical lines on either side of the peak maxima mark the signal region based on the locations of 20\% of the peak height.}
  \label{fig:smearing_fit}
\end{figure}

The method smears $p$, $\theta$, and $\phi$ of the reconstructed charged particles, including electrons, protons, and pions using Gaussian functions determined through an iterative approach to match the $\pi^-$ MMSQ resolution of MC with the experiment over its full kinematic range. The smearing factors take into account the different resolution functions of the FD and CD reconstruction. The smearing factors were derived using the exclusive $ep \to e'\pi^+\pi^-p'$ topology. Figure~\ref{fig:smearing_fit} shows for FD protons of a representative momentum bin [2.4:2.6]~GeV, the $\pi^-$ MMSQ distribution before smearing (top), after smearing (middle), and the corresponding measured MMSQ distribution (bottom).

\subsection{Radiative Corrections}
\label{sec:rad-corr}

The incoming, scattering, and outgoing electrons can emit photons as they pass through matter, reducing their energy and, hence, reconstructing to a different $Q^2$ and $W$ compared to the true value at the interaction vertex. Radiative correction factors were applied to the cross sections to account for these radiative effects. The Mo and Tsai method \cite{Mo:1968cg}, although developed for inclusive electron scattering, is commonly used successfully for radiative corrections of two-pion electroproduction data~\cite{Fedotov:2008aa,CLAS:2017fja,CLAS:2018fon,CLAS:2023mfc,Trivedi:2018rgo}. 
In the TWOPEG implementation, a minimum photon-energy cutoff of $10$~MeV is used to separate the soft-photon contribution from the hard-photon contributions in the Mo and Tsai calculation. This cutoff is chosen to be below the experimental resolution, so the calculated radiative correction is not expected to depend significantly on its precise value.

\begin{figure}[!ht]
  \centering
  \includegraphics[width=1.0\columnwidth]{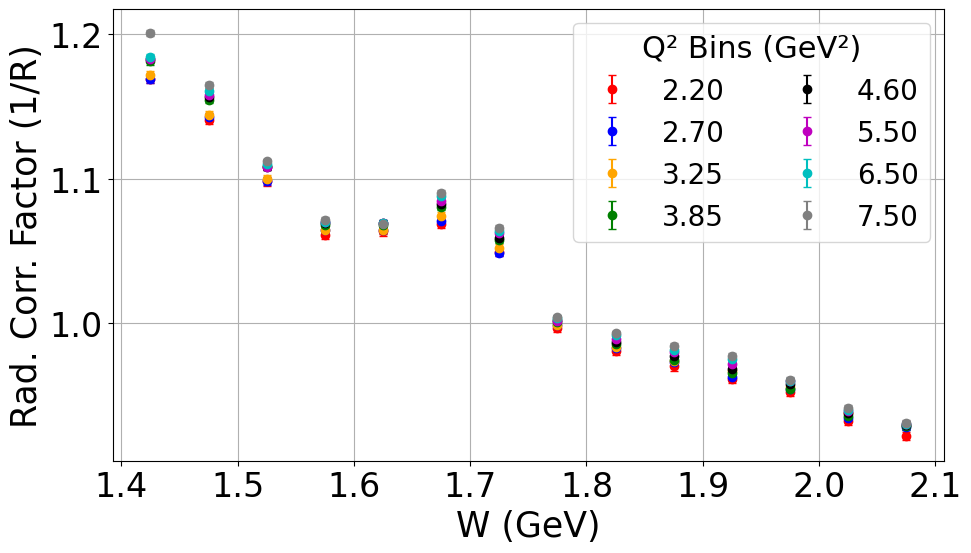}
  \caption{Radiative correction factors $1/R$ for various $Q^2$ bins in the $W$ range $[1.4, 2.1]$~GeV.}
  \label{fig:rad_corr_fact}
\end{figure}

The TWOPEG sample used to determine the acceptance factor $A$ was generated with radiative effects, including internal radiative effects and target straggling, and it was passed through the full GEMC simulation and reconstruction of the detector. Using radiative events for the acceptance calculation ensures that the simulated sample contains the same radiative kinematic migration as the experimental data. The radiative effect correction factors used in Eq.~(\ref{equ:cs-formula}) as $1/R$ is defined by
\begin{equation}
\frac{1}{R} = \frac{d^7\sigma_{\rm nr}(Q^2, W, \tau^5)}{d^7\sigma_{\rm r}(Q^2, W, \tau^5)}.
\label{equ:rad_corr_equ}
\end{equation}
Here, $d^7\sigma_{\rm nr}$ and $d^7\sigma_{\rm r}$ are the seven-fold differential cross sections without and with radiative effects, respectively. The correction can be simplified to depend only on $W$ and $Q^2$, as the radiative correction factor depends primarily on the electron vertex. Since TWOPEG weights the generated events according to the modeled cross sections, the correction factor can be expressed by
\begin{equation}
\frac{1}{R} = \frac{\Delta^2\Sigma_{\rm nr}(\Delta Q^2,\Delta W)}{\Delta^2\Sigma_{\rm r}(\Delta Q^2,\Delta W)},
\label{equ:rad_corr_equ1}
\end{equation}
\noindent
where $\Delta^2\Sigma_{\rm nr}$ and $\Delta^2\Sigma_{\rm r}$ represent the sum of all cross section weights in the respective $(\Delta Q^2,\Delta W)$ bins for events without and with radiative effects. Figure~\ref{fig:rad_corr_fact} shows these correction factors $1/R$ in different $Q^2$ bins for the range of $W \in [1.4,2.1]$~GeV. These corrections are predominantly in the range of $\pm 10\%$.

\subsection{Acceptance Studies and Uncertainty Cuts} 
\label{acept_error_ana}

Each $(W,Q^2)$ bin contains a large number of 5D bins, requiring extensive MC simulation to populate them adequately. Typically more than $10^8$ events per $(W,Q^2)$ bin are required based on previous CLAS analyses~\cite{CLAS:2023mfc,Trivedi:2018rgo}. TWOPEG generates events equally across the entire $(W,Q^2)$ range and assigns weights based on the cross section at each kinematic point. This optimized process overall requires much less MC simulation time, but slightly more time to adequately populate bins with smaller-sized $Q^2$ bins, since the same number of five-fold differential bins was included for each $(W,Q^2)$ bin. Thus, smaller-sized $Q^2$ bins tend to have larger hole fractions (where the concept of an acceptance hole is detailed in Section~\ref{sec:hole-filling}), which in turn leads to increased systematic uncertainties.

Although the model cross sections used to fill the empty cells provide a good description of the data, a very conservative relative systematic uncertainty of $\pm$50\% was assigned exclusively to the portion of the cross section originating from zero-acceptance cell contributions. For this reason, it is very important to sufficiently populate all 5D bins with MC events to ensure meaningful cross sections. Extensive simulations were performed over the full kinematic range $W \in [1.35,2.15]$~GeV, $Q^2 \in [1.95,9.0]$~GeV$^2$. 

\begin{figure}[!ht]
  \centering
  \includegraphics[width=1.0\columnwidth]{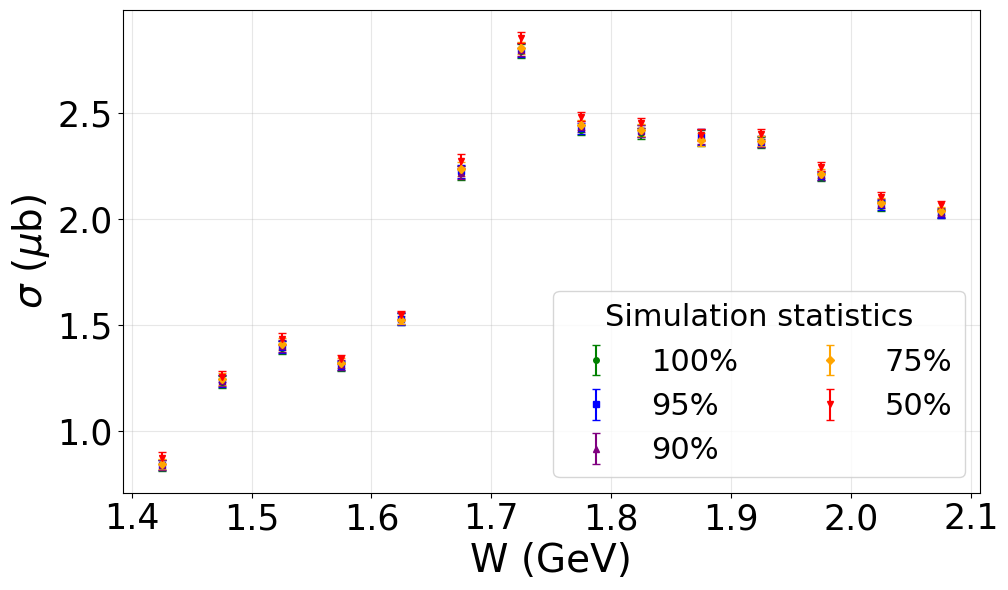}
  \caption{Integrated cross sections vs. $W$ extracted using different fractions of the available MC simulation statistics (50\%, 75\%, 90\%, 95\%, and 100\%) for the representative bin $Q^2 \in [3.5,4.2]$~GeV$^2$. The error bars represent the corresponding statistical uncertainties.}
  \label{fig:sim_stat_int}
\end{figure}

\begin{figure*}[!ht]
  \centering
  \includegraphics[width=0.80\textwidth]{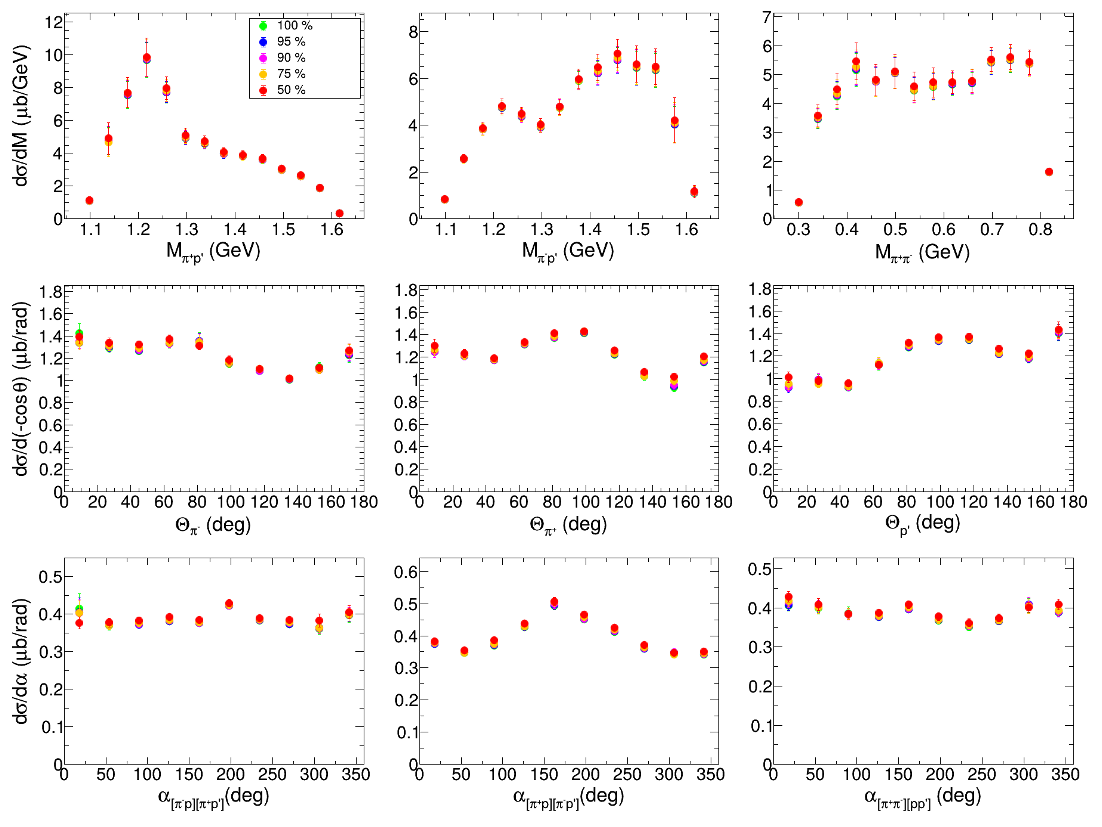}
  \caption{Nine one-fold differential cross sections extracted using different fractions of the available MC simulation statistics (50\%, 75\%, 90\%, 95\%, and 100\%) for a representative bin $W \in [1.75,\,1.80]$~GeV and $Q^2 \in [3.5,4.2]$~GeV$^2$.}
  \label{fig:sim_stat_1d}
\end{figure*}

The adequacy of the MC simulation statistics was evaluated by comparing the cross sections extracted using 50\%, 75\%, 90\%, 95\%, and 100\% of the available simulated events. Figures~\ref{fig:sim_stat_int} and~\ref{fig:sim_stat_1d} show the integrated and representative one-fold differential cross sections, respectively, for selected kinematic bins. As the simulation statistics increases, a larger fraction of the 7D kinematic bins become adequately populated, resulting in more reliable acceptance factor calculations and increased stability of the extracted cross sections. In contrast, with limited simulation statistics, bins with very small yields and poorly determined acceptances can contribute disproportionately, leading to an artificial enhancement of the extracted cross sections. To mitigate against such unphysical effects on the final cross sections, cuts were placed on the relative acceptance uncertainty $\delta A/A$ based on the studies in Refs.~\cite{Trivedi:ananote,Trivedi:2018rgo}.

\begin{figure}[!ht]
  \centering
  \includegraphics[width=0.95\columnwidth]{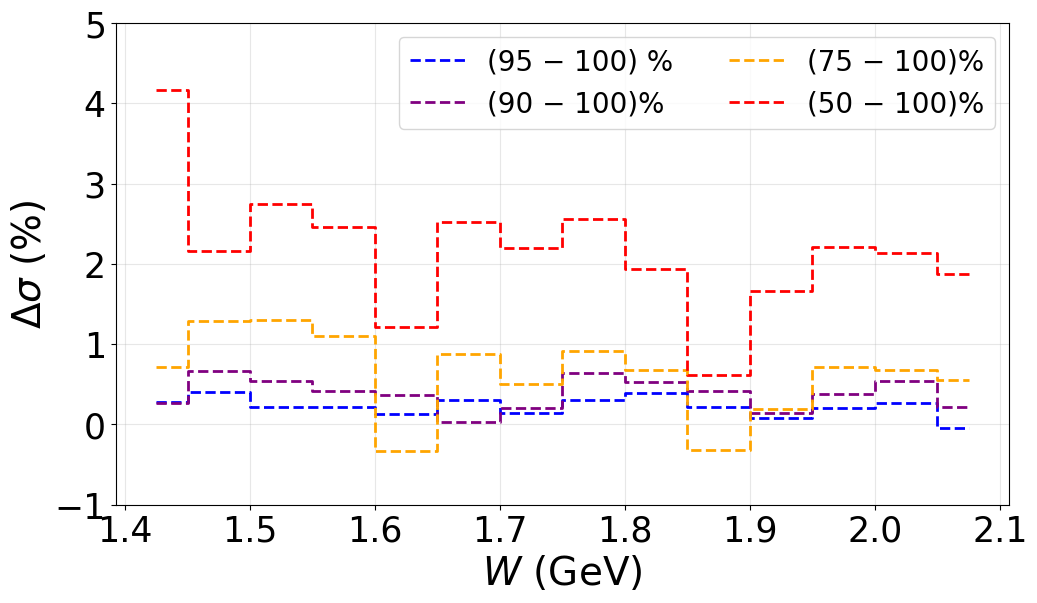}
  \caption{Difference (in \%) in the integrated cross sections extracted using different fractions of the available MC simulation statistics (50\%, 75\%, 90\%, 95\%, and 100\%) for $Q^2 \in [3.5,4.2]$~GeV$^2$.}
  \label{fig:diff_sim_stat_int}
\end{figure}

\begin{figure*}[!ht]
  \centering
  \includegraphics[width=0.70\textwidth]{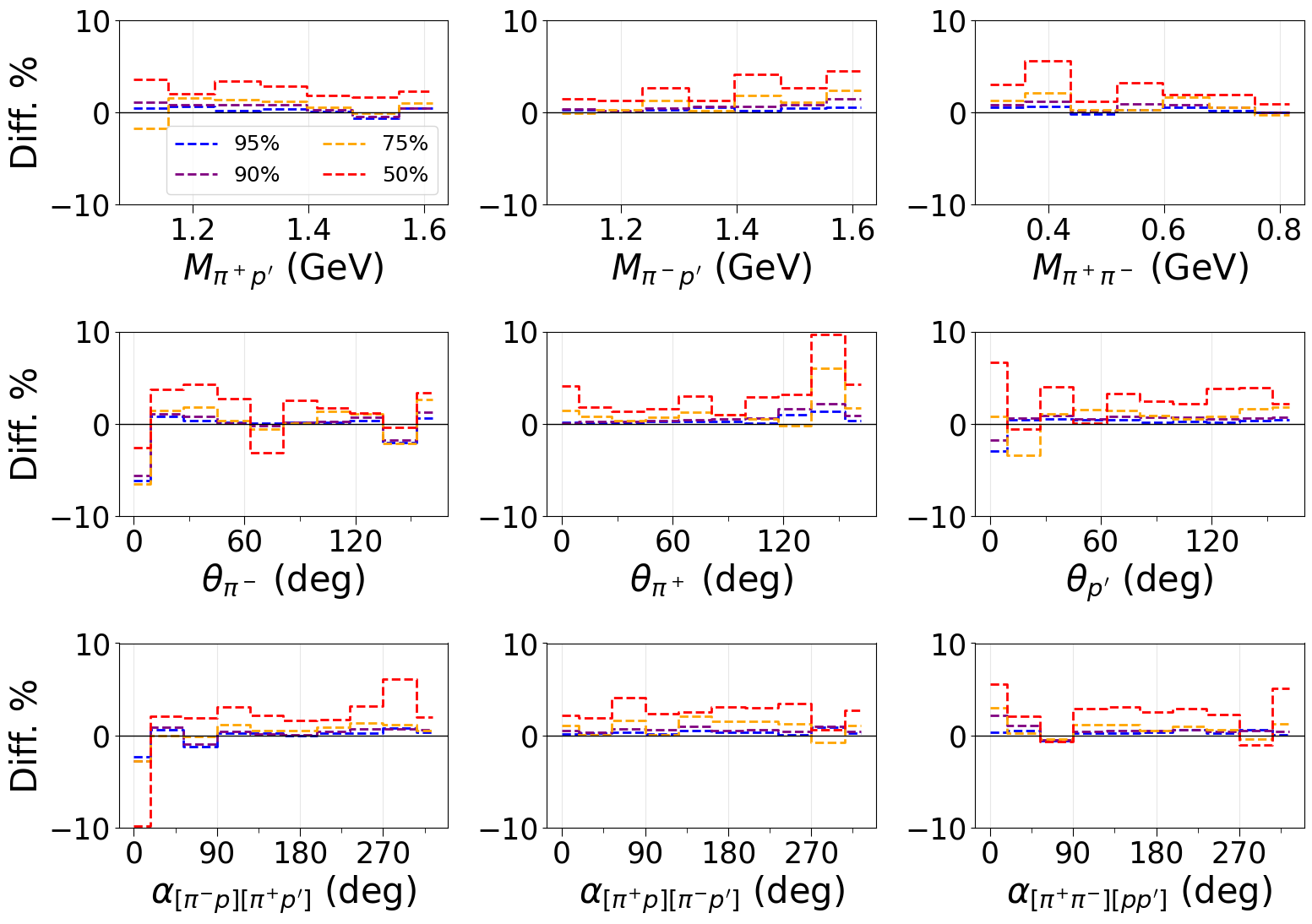}
  \caption{Difference (in \%) in the nine 1D differential cross sections extracted using different fractions of MC simulation statistics (50\%, 75\%, 90\%, 95\%, and 100\%) for $W \in [1.75,1.80]$~GeV and $Q^2 \in [3.5,4.2]$~GeV$^2$.}
  \label{fig:diff_sim_stat_1d}
\end{figure*}

Figure~\ref{fig:diff_sim_stat_int} shows the relative differences between the integrated cross sections obtained with the various simulation statistics. The differences between the cross sections obtained using 50\%, 75\%, and 100\% simulation data are found to be up to 5\% but converge for the difference between 95\% and 100\% to less than 0.5\%. A similar study was performed for the nine one-fold differential cross sections to assess the impact of limited MC simulation statistics on their stability. Figure~\ref{fig:diff_sim_stat_1d} shows the difference in the nine one-fold differential cross sections extracted using reduced simulation statistics relative to the full available simulation statistics (100\%). For the lowest simulation statistics (50\%), the observed differences are mostly within 5\% but reach up to $\approx 10\%$ in some edge bins, indicating a noticeable sensitivity to insufficiently populated kinematic bins and unstable acceptance. As the simulation statistics increases, these differences steadily decrease and the extracted differential cross sections exhibit a clear saturation behavior. In particular, for simulation statistics at the 90\% and 95\% levels, no significant systematic change in the extracted cross sections was observed. This behavior demonstrates that the available MC statistics are sufficient to ensure a stable and reliable extraction of the nine one-fold differential cross sections. 

\begin{figure}[ht]
  \centering
  \includegraphics[width=1.0\columnwidth]{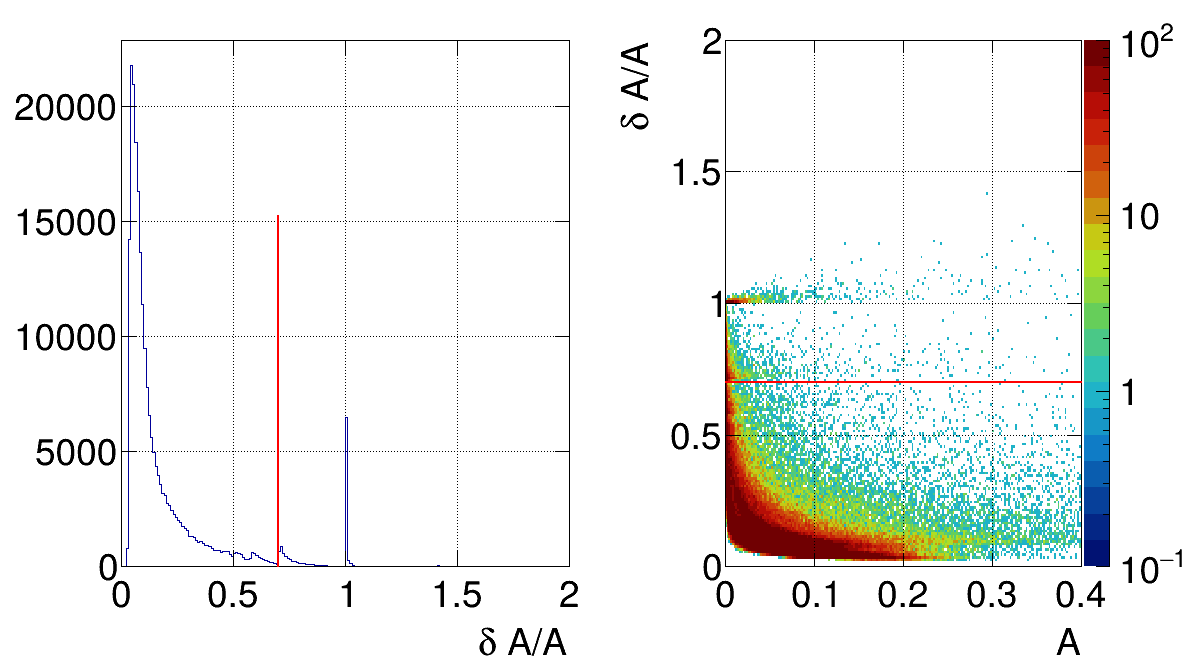}
  \caption{Relative acceptance uncertainty $\delta A/A$ (left) and relative acceptance uncertainty $\delta A/A$ vs. acceptance $A$ (right) for the representative bin $W \in [1.80, 1.85]$~GeV, $Q^2 \in [2.4, 3.0]$~GeV$^2$. The vertical red line in the left plot and the horizontal red line in the right plot represent the $\delta A/A$ threshold cut used.}
  \label{fig:eff_err_cut1}
\end{figure}

When the bin content of the acceptance histogram is small, the value of the relative acceptance uncertainty $\delta A/A$ is large and not well controlled. In such cases, the statistical fluctuations in the bin content become large relative to its value, leading to large relative acceptance uncertainties. Figure~\ref{fig:eff_err_cut1} illustrates the relative acceptance uncertainty cuts applied for a representative $(W,Q^2)$ bin. Any 5D bins with a $\delta A/A$ ratio larger than the chosen threshold were considered unreliable and were removed from the direct cross section extraction process by setting their acceptance factors to zero. These bins were thus treated as holes and later filled using TWOPEG model cross sections, as described in the next section. The acceptance uncertainty cuts were optimized separately for each $Q^2$ bin, since the statistical conditions vary with $Q^2$. Although at low $Q^2$ the experimental yields are high, the relevant MC statistics are relatively limited and hence the relative acceptance uncertainties were pushed to larger values, demanding a higher cut value to remove the worst uncertainties while keeping the hole-filling fraction in check. At higher $Q^2$ a tighter cut was applied to reduce the hole-filling systematic uncertainty further. The cut value of $\delta A/A$ varies in the range from 0.7 at low $Q^2$ to 0.3 at high $Q^2$. The assigned systematic uncertainty for this cut is detailed in Section~\ref{sec:syst}.

\subsection{Hole Filling}
\label{sec:hole-filling}

Certain five-fold differential bins in the simulation may not contain thrown or reconstructed two-pion events due to the resolution or limited detector acceptance, respectively. These bins with zero acceptance are called holes. As discussed in the previous section, the acceptance uncertainty cut creates more holes by setting the acceptance to zero. The nine one-fold differential cross sections were obtained by integrating the five-fold differential cross sections over different sets of the four hadronic variables. The evaluation of the nine one-fold differential cross sections, hence, requires that the 5-fold differential cross sections in zero-acceptance 5D cells have to be estimated and included into the integration employed for the nine one-fold differential cross section extraction. 

The reconstructed experimental and MC yields from all filled MC cells (see Eq.~(\ref{eq:cellyields})) were used to estimate the anticipated experimental yields in the holes, which are called hole yields. Hole-filling corrections on the extracted cross sections were done using those hole yields. The ratio of the sum of experimental over MC acceptance-corrected yields across all filled 5D MC cells determined the hole filling scaling factor ($ScF$) in each given $(W,Q^2)$ bin.

In more detail, the hole-filling procedure is described at the level of individual 5D bins in each given $(W,Q^2)$ bin. For a given 5D bin $i$, let $N^{\text{MCgen}}_i$, $N^{\text{MCrec}}_i$, and $N^{\text{EXPrec}}_i$ denote the generated MC, reconstructed MC, and reconstructed experimental yields, respectively. For the following one should keep in mind that the acceptance correction is implemented as a 5D-histogram division (ROOT \texttt{Divide}\cite{Brun:1996ki}), which by convention returns a bin content of exactly zero, rather than performing an undefined division, whenever the denominator bin is zero. The acceptance in each 5D bin $i$ is defined from MC as
\begin{equation}
\label{eq:acceptance}
A_i = \frac{N^{\mathrm{MCrec}}_i}{N^{\mathrm{MCgen}}_i}.
\end{equation}
The acceptance-corrected MC and experimental yields are given by
\begin{equation}
\label{eq:cellyields}
N^{\mathrm{MC}}_i = \frac{N^{\mathrm{MCrec}}_i}{A_i},
\qquad
N^{\mathrm{Exp}}_i = \frac{N^{\mathrm{EXPrec}}_i}{A_i}.
\end{equation}
The hole yield in the simulation for bin $i$ is then defined as
\begin{equation}
H^{\mathrm{MC}}_i = N^{\mathrm{MCgen}}_i - N^{\mathrm{MC}}_i.
\end{equation}
The global hole-filling scaling factor is defined as
\begin{equation}
ScF = \frac{\sum_{i=1}^{N} N^{\mathrm{Exp}}_i}{\sum_{i=1}^{N} N^{\mathrm{MC}}_i},
\label{equ:scf_def}
\end{equation}
where the sum runs over all filled MC 5D cells, $N^{\mathrm{MC}}_i\neq0$. The final hole-filled data yield in each bin was given by
\begin{equation}
N^{\mathrm{HF}}_i = N^{\mathrm{EXPrec}}_i + ScF \times H^{\mathrm{MC}}_i \times M\!F,
\label{equ:hole_filling}
\end{equation}
\noindent
where the additional correction factor $M\!F$ is included to account for differences between the TWOPEG model and the measured CLAS12 cross sections.

\begin{figure}[ht]
  \centering
  \includegraphics[width=0.9\columnwidth]{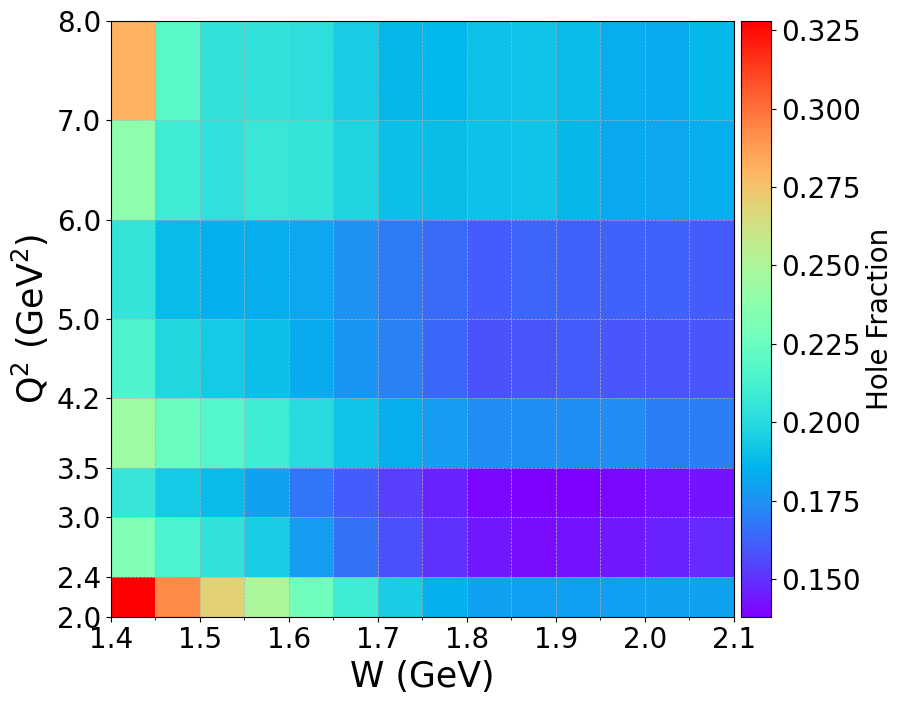}
  \caption{Fraction of holes (empty 5D bins) to all 5D bins as a function of $Q^2$ and $W$ from MC data.}
  \label{fig:hole_fractions}
\end{figure}

Figure~\ref{fig:hole_fractions} shows the fraction of empty 5D bins for each $(W,Q^2)$ bin after acceptance uncertainty cuts were applied. The hole fraction is higher near the two-pion production threshold in the low-$W$ region, where the cross sections are small. The figure also shows that the lower $Q^2$ region, independent of the fact that most experimental data are concentrated here, has as previously mentioned, lower simulation statistics and, hence, more holes. The lowest $Q^2$ bin has up to $\sim 30\%$ holes in its 5D bins. 

\subsection{Bin-Centering Corrections}
\label{sec:bin-centering}

The extracted single-differential and integrated cross sections are reported at the geometric center of each bin. However, the extracted values represent the average cross section across the entire bin. Therefore, bin-centering corrections were required to account for non-linear behavior of the cross sections within the kinematic bins. In this analysis, single-differential cross sections are reported in 3D bins in $W$, $Q^2$, and one of the hadronic variables. As a result, the bin-centering corrections must be calculated and applied to each bin in which the cross section is reported. The remaining four hadronic variables, being integrated over during the single-differential cross section extraction process, do not require bin-centering corrections.

\begin{figure*}[ht]
  \centering
  \includegraphics[width=0.75\textwidth]{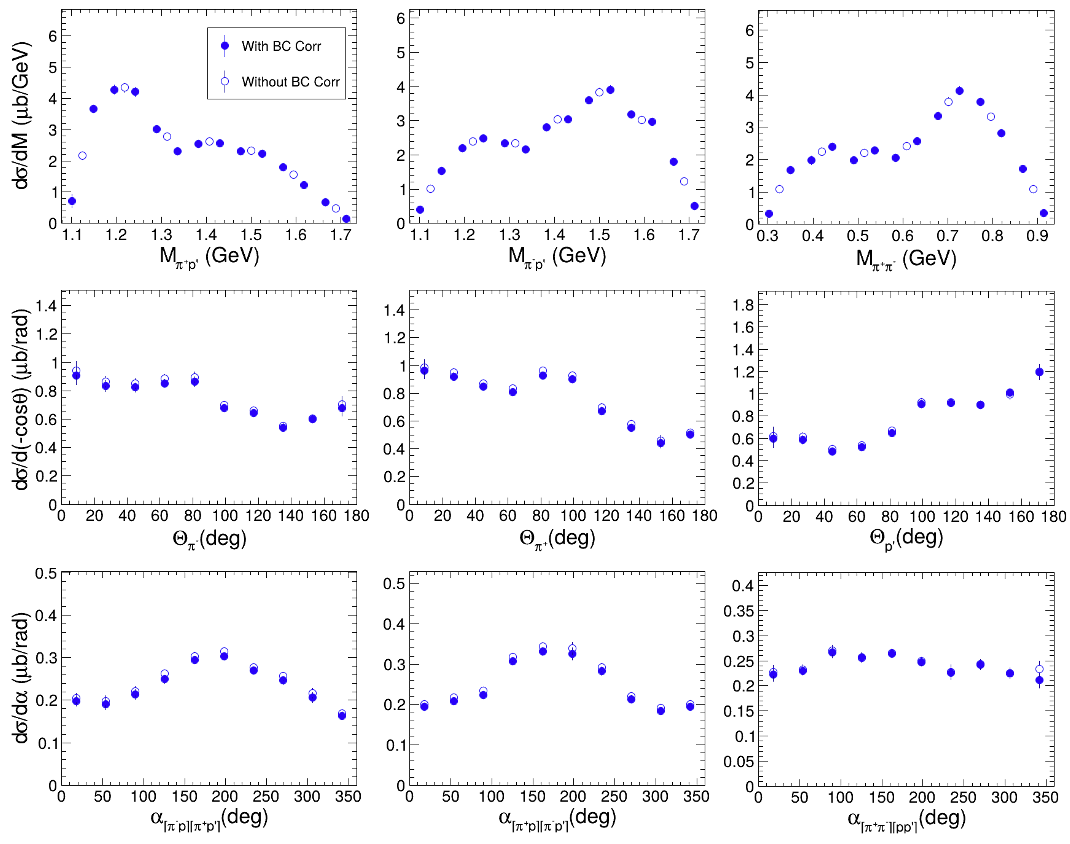}
  \caption{Effect of the bin-centering corrections for the nine one-fold differential cross sections for the representative bin $W \in [1.85, 1.90]$~GeV and $Q^2 \in [4.2, 5.0]$~GeV$^2$. The filled markers represent after the bin-centering corrections and the empty markers before. The error bars include only statistical uncertainties.}
  \label{fig:bin_centering_corr_1d}
\end{figure*}

The bin-centering correction factor was determined using event yields generated by the TWOPEG event generator. For each bin where single-differential cross sections are reported, the correction factor was calculated individually for $W$, $Q^2$, and the selected hadronic variable. The correction factor for a specific $W$ bin, $BC_{\text{corr}}(W)$, was determined subdividing the $W$ distribution in a given 3D bin into 11 smaller sub-bins. The correction factor was defined as the ratio of the yield in the central sub-bin to the average yield across all sub-bins. Similar factors were determined for the $Q^2$ bin centering correction factor, $BC_{\text{corr}}(Q^2)$, and the factor for centering within the hadronic variable, $BC_{\text{corr}}(\text{Hadronic Variable}$). As the original seven invariant-mass bins were divided into two equal half-bins to better represent the cross section variation, giving a total of 14 bins, the bin-centering correction was applied after this splitting. For each pair of half-bins, correction factors were calculated separately. The applied bin-centering correction factor for a given 3D bin, $BC_{\text{corr}}$, included in Eq.~(\ref{equ:cs-formula}) was defined as the product of the individual factors for $W$, $Q^2$, and the selected hadronic variable. Figure~\ref{fig:bin_centering_corr_1d} shows the effect of the bin-centering corrections on the nine one-fold differential cross sections for the representative bin $W \in [1.85, 1.90]$~GeV, $Q^2 \in [4.2, 5.0]$~GeV$^2$.

\section{Statistical and Systematic Uncertainties}
\label{sec:uncertainties}

\subsection{Statistical Uncertainties}

The statistical uncertainties in this analysis were evaluated separately for the experimental and MC yields, and were propagated to the final cross sections. For the experimental data, events were filled into 5D histograms with appropriate event weights for each $(W,Q^2)$ bin. The weight $w_i$ applied to each event corresponds to the reconstruction efficiency correction factors, as described in Section~\ref{sec:part_eff}. The statistical uncertainty in each bin was determined from the weighted event distribution

\begin{equation}
\delta N^{\cal E}_{\text{exp}}(\Delta^5\tau) = \sqrt{\sum_i w_i^2},
\end{equation}
\noindent
where $w_i$ is the weight of the $i^{th}$ event contributing to the bin. This corresponds to the standard expression for the variance of a weighted sum of independent events.

The treatment of the statistical MC uncertainties required special care due to the presence of cross section event weights from the TWOPEG event generator. For this purpose, the method described in Ref.~\cite{mc_stat_err} was used as in the previous charged two-pion analysis~\cite{CLAS:2023mfc,iuliia_th} in which the TWOPEG generator was used. This approach provided an analytic expression for the absolute statistical uncertainty of the acceptance in a given 5D bin for the case of weighted MC events as,
\begin{equation}
\delta {A}(\Delta^5\tau) =
\sqrt{
\frac{N_{\text{gen}} - 2N_{\text{rec}}}{N_{\text{gen}}^{3}} \sum_{i=1}^{N_{\text{rec}}} w_i^2 + \frac{N_{\text{rec}}^2}{N_{\text{gen}}^{4}} \sum_{j=1}^{N_{\text{gen}}} w_j^2}.
\label{eq:acceptance_error}
\end{equation}
Here, $w_i$ and $w_j$ correspond to the event-based TWOPEG cross section weights. This expression was taken as the statistical uncertainty of the acceptance in each $A\!>\!0$ bin.

Since the differential cross sections were extracted as the ratio of the fitted data yield to the acceptance, the final relative statistical uncertainty of the cross section for each 5D kinematic bin is given by
\begin{equation}
\frac{\delta \sigma(\Delta^5\tau)}{\sigma(\Delta^5\tau)} = \sqrt{\left( \frac{\delta N^{\cal E}_{\text{exp}}}{N^{\cal E}_{\text{exp}}} \right)^2 + \left( \frac{\delta {A}}{{A}} \right)^2 }.
\end{equation}
For the empty 5D bins, i.e. $A\!=\!0$, zero statistical uncertainty was assigned. Since the MC samples have significantly higher statistics than the measured data, the overall statistical uncertainty was generally dominated by the measured two-pion yields.

\subsection{Systematic Uncertainties} 
\label{sec:syst}

Each selection cut and correction applied in this analysis introduced a possible source of systematic uncertainty on the extracted cross sections. These uncertainties were evaluated by varying each selection criterion using loose, medium, and tight definitions in order to estimate their impact. For each source, the average deviation of the extracted cross section from the nominal (medium-cut) value was taken as the associated systematic uncertainty. The total systematic uncertainty was obtained by combining the individual contributions in quadrature. The overall average systematic uncertainty for the integrated cross sections was determined to be 9.0\%, as summarized in Table~\ref{tab:syst_unc}. The individual sources of systematic uncertainty in this analysis are detailed below.

\begin{table*}[htbp]
\centering
\caption{Average systematic uncertainties for the two-pion electroproduction cross sections over all $(W,Q^2)$ kinematic bins. See the text for details.}
\label{tab:syst_unc}
\begin{tabular}{|c|c|c|} \hline
Item & Source                          & Uncertainty (\%) \\ \hline
1    & PID, fiducial volume, MMSQ cuts & 2.0 \\ \hline
2    & Momentum corrections            & 0.8 \\ \hline
3    & Particle efficiency corrections & 2.8 \\ \hline
4    & Background merging              & 3.0 \\ \hline
5    & Smearing MC data                & 1.4 \\ \hline
6    & Background subtraction          & 0.1 \\ \hline
7    & Acceptance uncertainty cut      & 0.5 \\ \hline
8    & Radiative effects               & 5.0 \\ \hline
9    & Variable set dependence         & 1.6 \\ \hline
10   & Model dependence                & 1.4 \\ \hline
11   & Kinematic hole filling          & 4.6 \\ \hline
12   & Bin centering                   & 0.5 \\ \hline
13   & Target length                   & 1.0 \\ \hline
14   & HTCC efficiency                 & 2.0 \\ \hline
15   & Accumulated FC charge           & 1.2 \\ \hline \hline
     & Total uncertainty               & 9.0 \\ \hline
\end{tabular}
\end{table*}

\vskip 0.5cm

\noindent
1) \underline{Event selection cuts}: Each of the cuts associated with the two-pion event selection was studied individually. These cuts include the ECAL sampling fraction cut, PCAL energy cut, particle vertex cuts, fiducial volume cuts for the FD and CD, hadron PID $\Delta t$ cuts, and $\pi^-$ MMSQ cuts. Each cut was varied over a broad range to define the loose and tight cut versions and contributed to the systematic uncertainty at the level of $\lesssim 1\%$. The quadrature sum of these individual sources was conservatively assigned as 2\%.

\noindent
2) \underline{Momentum corrections}: The systematic uncertainty associated with the electron and hadron momentum corrections was evaluated by varying the applied correction factors by $\pm 10\%$. The resulting combined effect corresponds to a systematic uncertainty of $0.8\%$ on the integrated cross sections.

\noindent
3) \underline{Efficiency corrections}: For the particle reconstruction efficiency corrections ($\cal{E}$), three independent sources of systematic uncertainty were considered. First, the MMSQ selection boundaries were varied by $\pm 0.5\sigma$ obtained from Gaussian fits to the signal peaks. Second, the normalization of the exclusive background contribution was varied by $\pm 10\%$ to account for uncertainties in the background scaling procedure. Third, the uncertainty of the background shape was evaluated by varying the MMSQ selection criteria for the different reaction topologies, including the missing proton, missing $\pi^+$, and fully exclusive channels, which determine the background shape. The combined contribution of these sources resulted in a systematic uncertainty of $\approx 2.8\%$ in the integrated cross sections.

\noindent
4) \underline{Background merging}: The systematic uncertainty associated with background merging was estimated following the methodology of Ref.~\cite{CLAS:2025zup}. The background merging used background files for this dataset at the three different beam currents. A systematic uncertainty of 3\% was assigned associated with the difference between the experiment and MC derived tracking efficiencies for the electron.

\noindent
5) \underline{Smearing of MC data}: The systematic uncertainty associated with the smearing procedure was evaluated by varying the smearing factors for the electron and hadrons $p$, $\theta$, and $\phi$ by $\pm 10\%$ relative to their nominal values. This source contributes 1.4\% to the systematic uncertainty on the integrated cross sections.

\noindent
6) \underline{Background subtraction}: To evaluate the systematic uncertainty associated with background subtraction, an alternative extrapolation procedure towards low $W$ was performed to mitigate the larger uncertainties at smaller invariant masses, in which the physical production energy threshold of three-pion production was explicitly enforced. The difference between the two background-subtraction scenarios resulted in an average variation of 0.1\% for the integrated cross section.

\noindent
7) \underline{Acceptance-uncertainty cuts}: The systematic uncertainty associated with the relative acceptance-uncertainty cuts was evaluated by varying the relative acceptance-uncertainty thresholds around their nominal values over an appropriate data-driven range. The resulting variation in the integrated cross sections contributed at the level of 0.5\%.

\noindent
8) \underline{Radiative effects}: The typical systematic uncertainty associated with the radiative corrections in the two-pion electroproduction channel was taken from Ref.~\cite{CLAS:2017fja} where a 5\% scale uncertainty was assigned.

\noindent
9) \underline{Variable set dependence}: The systematic uncertainty associated with the choice of the set of 5D hadronic variables was evaluated by extracting the integrated cross section using each of the variable sets described in Section~\ref{sec:formalism}. For each $(W,Q^2)$ bin, an integrated cross section was obtained independently for each variable set by integrating the corresponding one-fold differential invariant-mass cross sections, namely with respect to $M_{\pi^+p'}$, $M_{\pi^-p'}$, and $M_{\pi^+\pi^-}$. The final value was taken as the average of the results from the three sets, while the spread among them was taken as the systematic uncertainty due to the variable-set dependence, following the prescription of Refs.~\cite{CLAS:2023mfc,iuliia_th}. This source contributed an average systematic uncertainty of $1.6\%$ to the integrated cross sections.

\noindent
10) \underline{Model dependence}: This source was estimated by comparing the cross sections obtained using the original TWOPEG generator with those obtained after applying the $Q^2$-dependent and $(W,Q^2)$ bin-by-bin scaling adjustments described in Section~\ref{sec:cs_scaling_factor}. Because the modeled cross sections were directly used in the hole-filling procedure, the impact of the adjusted TWOPEG cross sections affected the hole-filled regions of the phase space. The difference of 1.4\% between the resulting extracted integrated cross sections was assigned as the systematic uncertainty.

\noindent
11) \underline{Kinematic hole-filling}: In regions of phase space where the detector acceptance is zero, the corresponding cross section values cannot be extracted directly from the experimental data. These kinematic holes were filled using cross sections generated by the TWOPEG event generator. As a result, the hole-filling correction is model dependent. Despite the small impact of the model dependence noted in item \#10, we accounted for this effect by assigning a very conservative systematic uncertainty of 50\% of the estimated cross section to each hole filled. The overall systematic uncertainty associated with this source was estimated to be 4.6\% for the integrated cross sections.

\noindent
12) \underline{Bin centering}: The average statistical uncertainty associated with the bin-centering corrections was found to be 0.5\%. This value was assigned as the systematic uncertainty due to the bin-centering correction and included in the total systematic uncertainty of the measured cross sections.

\noindent
13) \underline{Target length}: The nominal length of the cryotarget cell is 5.0~cm with a mechanical tolerance of $\pm 0.05$~cm. This give rise to a scale uncertainty on the cross sections of 1\%.

\noindent
14) \underline{HTCC Efficiency}: A study of the HTCC efficiency was performed and reported in the CLAS12 inclusive analysis~\cite{CLAS:2025zup}, where the efficiency was determined using measured and MC data. The observed differences were attributed to mirror non-uniformities and geometrical effects. Since no dedicated study was performed and no explicit HTCC efficiency correction was applied in this analysis, a conservative scale systematic uncertainty of 2\% was assigned to the final cross sections.

\noindent
15) \underline{Accumulated FC Charge}: The integrated luminosity was determined from the live-gated Faraday cup charge. Following the cross section analysis of Ref.~\cite{CLAS:2025zup}, performed with the same CLAS12 dataset, a total scale systematic uncertainty of 1.2\% was assigned to the accumulated FC charge.
\section{Results and Discussion}
\label{sec:results}

The nine one-fold differential and fully integrated $\pi^+ \pi^- p$ electroproduction cross sections have become available across the kinematic region $W \in [1.40,2.10]$~GeV and $Q^2 \in [2.4,8.0]$~GeV$^2$. The extracted cross sections over the full kinematic phase space are included in the CLAS Physics Database~\cite{clasphysdb}. The CLAS12 results reported in this paper provide, for the first time, coverage of the $Q^2 \in [5.0,8.0]$~GeV$^2$ range. The previously available CLAS data on the nine one-fold differential and integrated $\pi^+ \pi^- p$ cross sections \cite{Trivedi:2018rgo,CLAS:2017fja} have substantial overlap with the present CLAS12 measurements across $W \in [1.40,2.00]$~GeV for $Q^2 \in [2.0,5.0]$~GeV$^2$.

It is important to note that the previous CLAS results were obtained with finer $W$ bin sizes of 25~MeV, whereas this analysis uses 50~MeV bins. Therefore, CLAS has twice as many $W$ bins for the cross sections as CLAS12. In the comparison of the single-differential CLAS and CLAS12 cross sections, the results from CLAS12 were thus compared with those averaged over the two $W$ bins measured with CLAS that lie within a single $W$ bin measured with CLAS12. Furthermore, in contrast to the earlier CLAS measurements, the CLAS12 analysis employs bin-centering corrections. The knowledge of the shapes of both the differential and integrated cross section data obtained with CLAS12 is thus improved relative to that obtained with CLAS.

\begin{figure*}[!ht]
\centering
\includegraphics[width=0.75\textwidth]{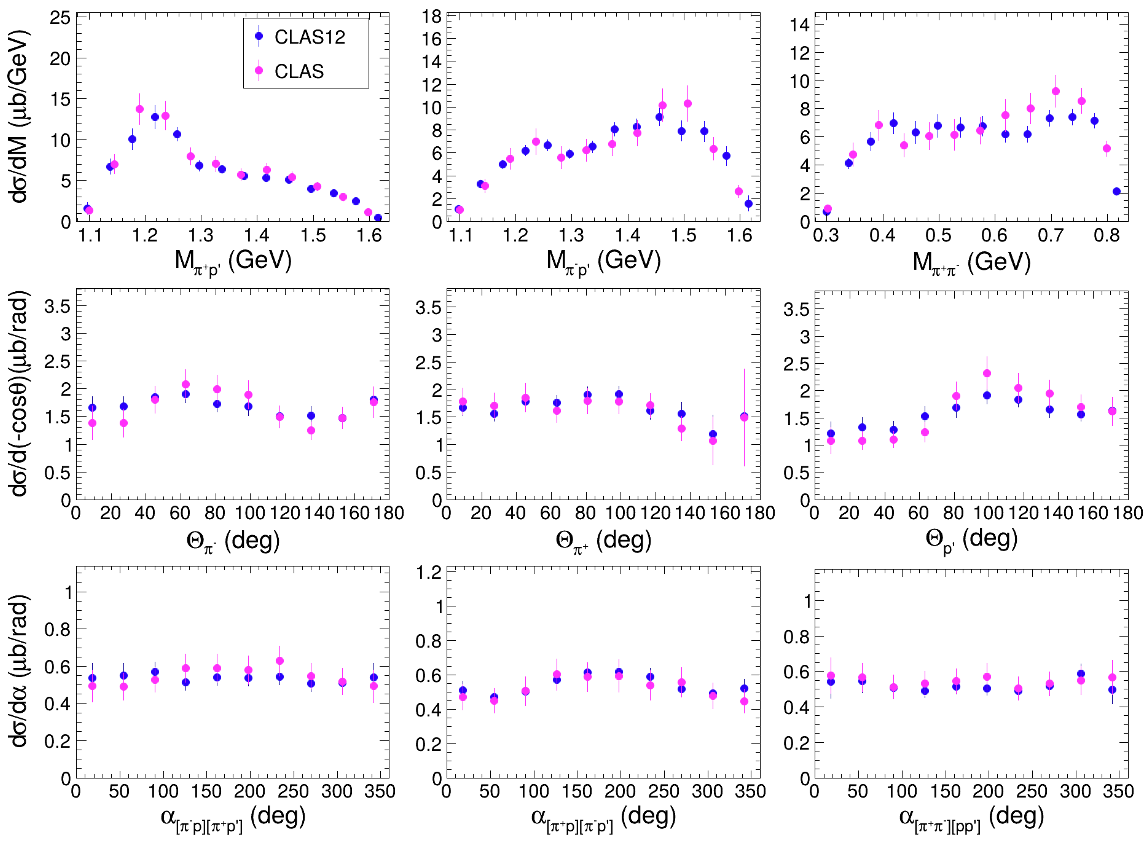}
\caption{Measured nine single-differential cross sections for $W \in [1.75,1.80]$~GeV and $Q^2 \in [3.0,3.5]$~GeV$^2$. The blue points with error bars are the CLAS12 cross sections, and the magenta points are the previous CLAS results~\cite{Trivedi:2018rgo}. The CLAS12 and CLAS data were obtained at different beam energies for different bin $W$ sizes, and bin-centering corrections were employed only in the CLAS12 measurements. For both measurements the uncertainties are the quadratic sum of their statistical and systematic uncertainties. }
\label{fig:diff_cs_vs_clas6}
\end{figure*}

\begin{figure*}[!ht]
\centering
\includegraphics[width=0.75\textwidth]{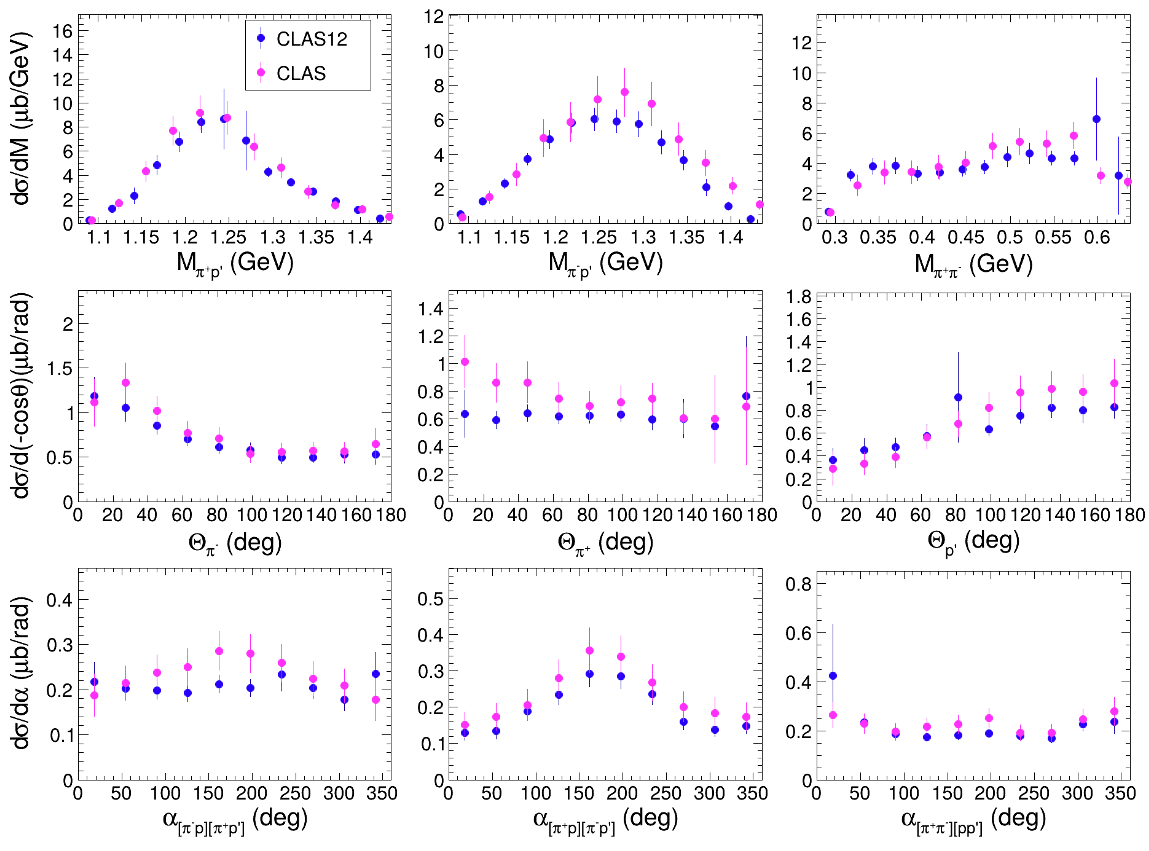}
\caption{Measured nine single-differential cross sections for $W \in [1.55,1.60]$~GeV and $Q^2 \in [3.5,4.2]$~GeV$^2$. The blue points with error bars are the CLAS12 cross sections, and the magenta points are the previous CLAS results~\cite{Trivedi:2018rgo}. The CLAS12 and CLAS data were obtained at different beam energies for different bin $W$ sizes, and bin-centering corrections were employed only in the CLAS12 measurements. For both measurements the uncertainties are the quadratic sum of their statistical and systematic uncertainties.}
\label{fig:diff1_cs_vs_clas6}
\end{figure*}

\begin{figure}[!ht]
\centering
\includegraphics[width=0.9\columnwidth]{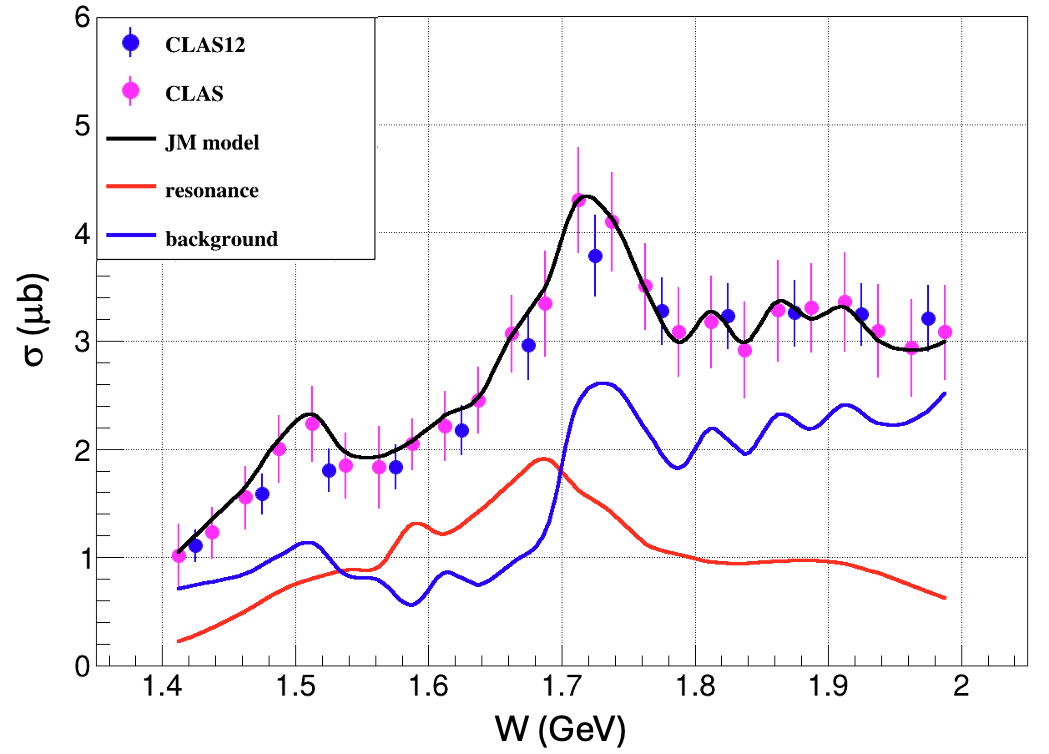}
\caption{Comparison of the measured integrated cross section versus $W$ for $Q^2 \in [3.0,3.5]$~GeV$^2$. The blue points are the CLAS12 cross sections and the magenta points are the previous CLAS results~\cite{Trivedi:2018rgo}. The error bars include the quadratic sum of the statistical and systematic uncertainties. The curves show the JM model calculations, with the total contribution (black), the resonant part (red), and the non-resonant background (blue).}
\label{fig:integrated_cs_vs_clas6}
\end{figure}

The differential and integrated cross sections measured with CLAS and CLAS12 agree with each other across the entire overlapping $(W,Q^2)$ range within their uncertainties for majority of the data points, showing consistent behavior in the kinematic dependence of these observables. Representative examples of the differential and fully integrated $\pi^+ \pi^- p$ electroproduction cross sections are shown in Figs.~\ref{fig:diff_cs_vs_clas6},~\ref{fig:diff1_cs_vs_clas6}, and~\ref{fig:integrated_cs_vs_clas6}. Remaining systematic differences between the differential cross sections measured for CLAS and CLAS12 may arise due to:
\begin{itemize}
    \item different electron beam energies;
    \item a factor of two broader bin size over $W$ for the CLAS12 dataset;
    \item bin-centering corrections were employed only for the CLAS12 data.
\end{itemize}

Analyses of the nine one-fold differential $\pi^+ \pi^- p$ electroproduction cross sections from CLAS data within the framework of the JM23 model version \cite{Mokeev:2023zhq} allowed all essential contributing mechanisms to be established through their manifestations in the observables, such as peaks in the invariant-mass distributions and pronounced evolutions in the $\theta$ and $\alpha$ angular distributions. These contributing mechanisms were identified by examining correlations between their characteristic shapes across the different one-fold differential cross sections \cite{Mokeev:2023zhq,2pi-hilevel}.

The good description of the one-fold differential cross sections measured with CLAS, achieved across the broad kinematic region of $W \in [1.4,2.0]$~GeV and $Q^2 \in [2.0,5.0]$~GeV$^2$, enables a reliable separation of the resonant and non-resonant contributions, which is crucial for the extraction of the $\gamma_v p N^*$ electrocouplings. A representative example of the separation between the resonant and non-resonant contributions to the fully integrated $\pi^+ \pi^- p$ electroproduction cross section is shown in Fig.~\ref{fig:integrated_cs_vs_clas6} illustrating its applicability to both the CLAS and CLAS12 datasets. The quality of the CLAS data has made it possible to determine the $\gamma_v p N^*$ electrocouplings for most $N^*$ states across the mass range up to 1.8~GeV for $Q^2 \in [2.0,5.0]$~GeV$^2$~\cite{Mokeev:2023zhq}. The extraction of the electrocouplings for $N^*$ in the mass range from 1.8 to 2.0~GeV is in progress, which will be followed by a high-level analysis of the new CLAS12 results presented here.

\begin{figure*}[!ht]
\centering
\includegraphics[width=0.75\textwidth]{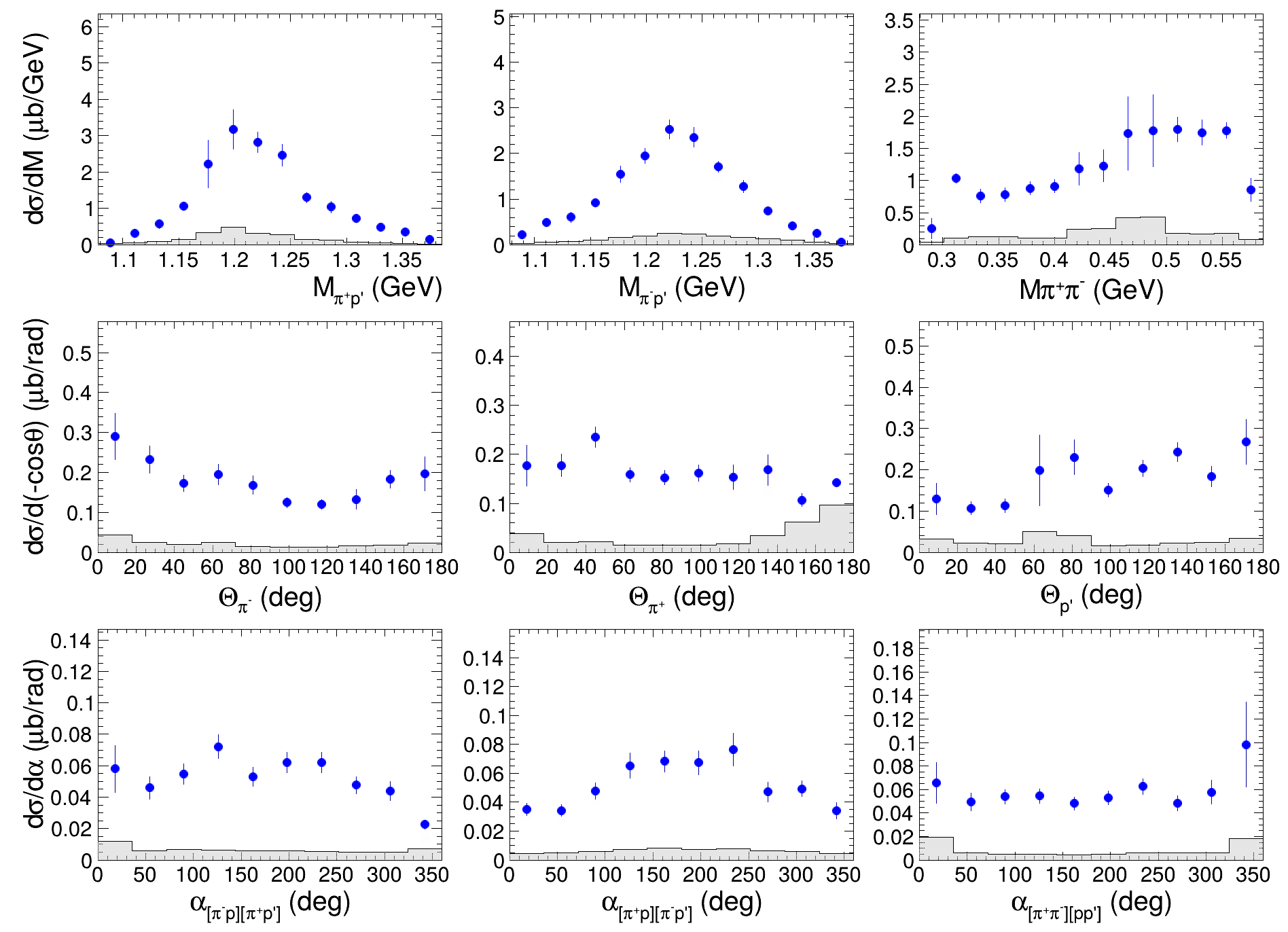}
\caption{Nine single-differential cross sections for $W \in [1.50,1.55]$~GeV and $Q^2 \in [6.0,7.0]$~GeV$^2$ measured with CLAS12. The error bars include only the statistical uncertainties. The filled gray histograms represent the systematic uncertainty. The full set of single-differential cross sections over the full measured kinematic phase space are included in the CLAS Physics Database~\cite{clasphysdb}.}
\label{fig:diff_cs_high_q2}
\end{figure*}

\begin{figure}[!ht]
\centering
\includegraphics[width=1.0\columnwidth]{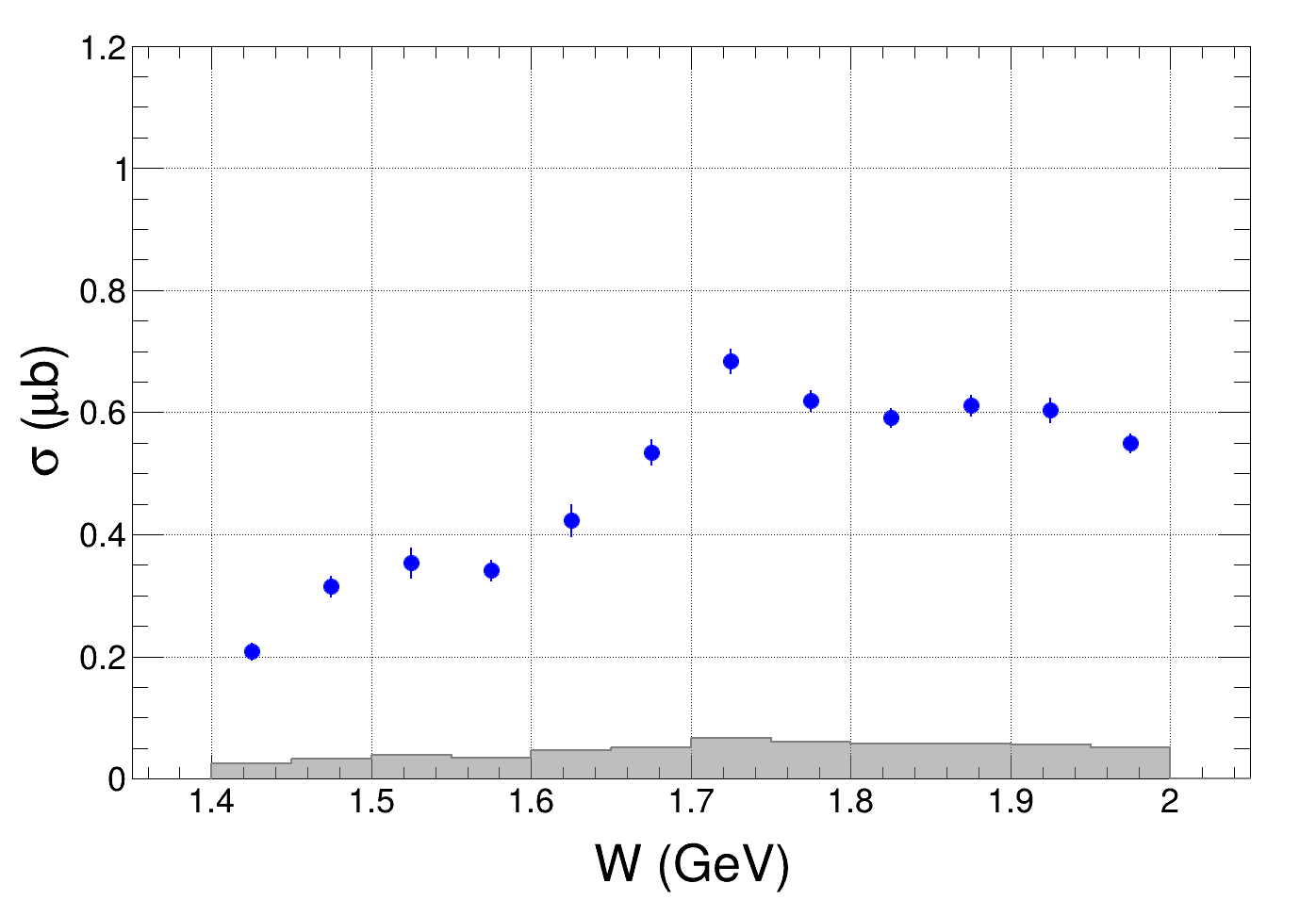}
\caption{Integrated measured cross section versus $W$ for $Q^2 \in [6.0,7.0]$~GeV$^2$ measured with CLAS12. The error bars include only the statistical uncertainties. The filled gray histograms represent the systematic uncertainties. The full set of integrated cross sections over the full measured kinematic phase space are included in the CLAS Physics Database~\cite{clasphysdb}.}
\label{fig:integrated_cs_high_q2}
\end{figure}

The differential and fully integrated $\pi^+ \pi^- p$ electroproduction cross sections reported in this work at $Q^2 = [5.0,8.0]$~GeV$^2$ have been obtained for the first time. Representative examples of the single-differential and integrated cross sections are shown in Figs.~\ref{fig:diff_cs_high_q2} and~\ref{fig:integrated_cs_high_q2}. These experimental data will allow for extension of the applicability of the JM model toward $Q^2 = 8.0$~GeV$^2$ across the $W < 2.0$~GeV range. 

The data quality, in terms of kinematic coverage and uncertainties, is comparable to that achieved with CLAS and previously used for the successful extraction of $\gamma_v p N^*$ electrocouplings within the JM model~\cite{Mokeev:2023zhq}. This offers strong prospects for extending the results on the $Q^2$ evolution of the resonance electrocouplings to $Q^2 = 8.0$~GeV$^2$ for all prominent excited states of the nucleon. The anticipated extension of the results on the $Q^2$ evolution of the $\gamma_vpN^*$ electrocouplings towards $Q^2$=8.0~GeV$^2$ will allow us to explore the range of distances where around 50\% of hadron mass emerges in the transition from strongly coupled to the perturbative regimes~\cite{Achenbach:2025kfx, Ding:2022ows}.

\section{Summary and Conclusions}
\label{sec:conclusions}

This analysis of CLAS12 data presents new integrated and nine one-fold differential cross sections for $\pi^+\pi^-p$ electroproduction over the nucleon resonance region for invariant masses $W$ from 1.4 to 2.1~GeV. The analysis was based on the CLAS12 RG-A fall 2018 dataset collected at a beam energy of 10.6~GeV and with nominal torus field set to inbending polarity, so that negatively charged particles bend toward the electron beamline. These data span a broader kinematic range, covering photon virtualities $Q^2$ from 2.4 to 8.0~GeV$^2$, than previous CLAS results that extend only up to 5.0~GeV$^2$. In each $(W,Q^2)$ bin, nine independent one-fold differential $\pi^+ \pi^- p$ electroproduction cross sections were extracted, which represent integrals of a common five-fold differential cross section over different sets of four kinematic variables of the final state hadrons. Across the overlapping region of $W < 2.0$~GeV and $Q^2 < 5.0$~GeV$^2$, the $\pi^+ \pi^- p$ electroproduction cross sections measured in the 6-GeV era with CLAS are consistent with the CLAS12 results reported here. The CLAS12 $\pi^+\pi^-p$ cross sections extracted for $Q^2$ in the range from 5.0 to 8.0~GeV$^2$ extend into previously unexplored kinematic regions where no prior measurements exist for this exclusive channel. 

Analyses of the CLAS data within the JM meson-baryon reaction model have provided information on nucleon resonance electrocouplings across the $Q^2$ range up to 5.0~GeV$^2$ \cite{Mokeev:2023zhq,2pi-hilevel}. Owing to the extended kinematic range of the CLAS12 data compared with earlier CLAS measurements, it is anticipated that the analysis of these new CLAS12 data within reaction models, in particular the JM model, will provide information on the $Q^2$ evolution of the $\gamma_v p N^*$ electrocouplings for all prominent nucleon resonances beyond the first resonance region over the mass range up to 2.0~GeV for $Q^2 < 8.0$~GeV$^2$. This $Q^2$ range corresponds to a distance scale where the contributions from dressed quarks to $N^*$ structure are expected to be dominant~\cite{Burkert:2019bhp,Achenbach:2025kfx}. The results on the $\gamma_v p N^*$ electrocouplings for $Q^2$ up to 8.0~GeV$^2$ are expected to be of particular importance when exploring the emergence of the strong-interaction dynamics and the underlying $N^*$ three-quark structure in the transition from the strongly coupled towards the perturbative QCD regimes.

\begin{acknowledgments}

The authors would like to acknowledge the outstanding efforts of the JLab staff that made this experiment possible. This work was supported in part by the National Science Foundation (NSF) under grants PHY 10011349 and 10014377, the U.S. Department of Energy (DOE), Office of Science, Office of Nuclear Physics under contract 89243126CSC000213. This work was furthermore supported in part by the University of South Carolina and the Center for Nuclear Femtography, operated by the Southeastern Universities Research Association in Washington, D.C. under an appropriation from the Commonwealth of Virginia, the Chilean Comisi\'on Nacional de Investigaci\'on Cient\'ifica y Tecnol\'ogica (CONICYT), the Italian Istituto Nazionale di Fisica Nucleare, the French Centre National de la Recherche Scientifique, the French Commissariat \`{a} l'Energie Atomique, the Scottish Universities Physics Alliance (SUPA), the United Kingdom's Science and Technology Facilities Council, the National Research Foundation of Korea, and the Skobeltsyn Nuclear Physics Institute and Physics Department at the Lomonosov Moscow State University. 

\end{acknowledgments}

\bibliography{references}{}
\bibliographystyle{apsrev4-1}
\end{document}